\documentclass[journal=jctcce,manuscript=article]{achemso}

\usepackage[version=3]{mhchem} 
\usepackage{graphicx,color}
\usepackage{bm}
\usepackage{hyperref}
\definecolor{myblue}{rgb}{0,0,1}
\hypersetup{
    unicode=false,          
    pdftoolbar=true,        
    pdfmenubar=true,        
    pdffitwindow=false,     
    pdfstartview={FitH},    
    pdfauthor={},         
    colorlinks=true,        
    linkcolor=myblue,         
    citecolor=myblue,          
}

\author{Chia-Nan Yeh}
\affiliation[CCQ]{Center for Computational Quantum Physics, Flatiron Institute, New York, New York 10010, USA}
\email{cyeh@flatironinstitute.org}
\author{Miguel A. Morales}
\affiliation[CCQ]{Center for Computational Quantum Physics, Flatiron Institute, New York, New York 10010, USA}
\email{mmorales@flatironinstitute.org}

\title{Low-Scaling Algorithm for the Random Phase Approximation using Tensor Hypercontraction with k-point Sampling}

\begin{document}

\begin{tocentry}
\includegraphics[width=0.9\textwidth]{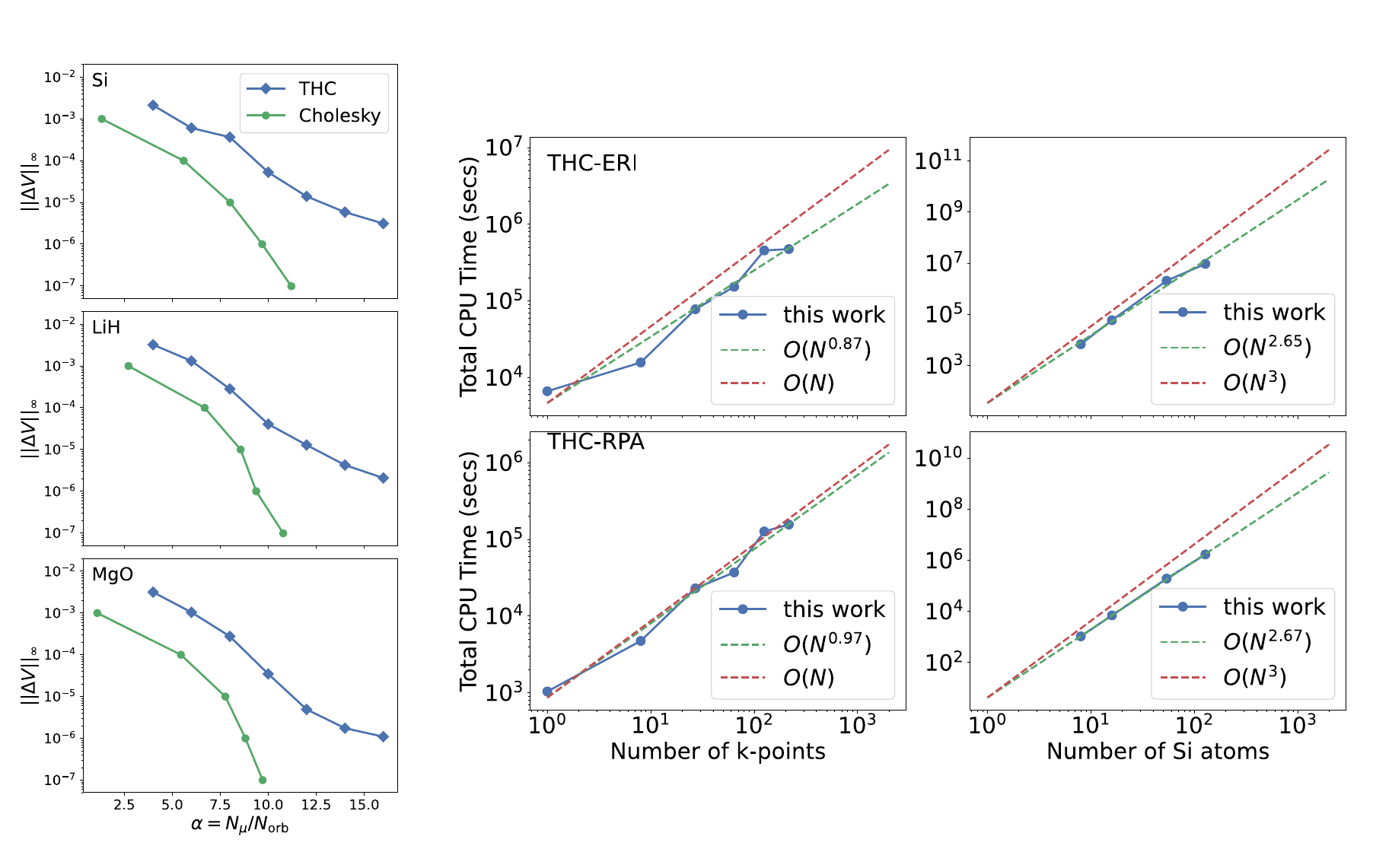}
\end{tocentry}

\begin{abstract}
  We present a low-scaling algorithm for the random phase approximation (RPA) with \textbf{k}-point sampling in the framework of tensor hypercontraction (THC) for electron repulsion integrals (ERIs). The THC factorization is obtained via a revised interpolative separable density fitting (ISDF) procedure with a momentum-dependent auxiliary basis for generic single-particle Bloch orbitals. Our formulation does not require pre-optimized interpolating points nor auxiliary bases, and the accuracy is systematically controlled by the number of interpolating points. The resulting RPA algorithm scales linearly with the number of \textbf{k}-points and cubically with the system size without any assumption on sparsity or locality of orbitals. The errors of ERIs and RPA energy show rapid convergence with respect to the size of the THC auxiliary basis, suggesting a promising and robust direction to construct efficient algorithms of higher-order many-body perturbation theories for large-scale systems. 
\end{abstract}

\section{Introduction\label{sec:intro}}
Kohn-Sham density functional theory (KS-DFT)\cite{Hohenberg1964,KSDFT_Sham1965} has become the standard tool in the study of ground-state properties of molecules and solids due to its capability of efficiently treating large-scale systems with reasonable accuracy. 
Nevertheless, there are many well-known cases in which DFT fails to provide even qualitatively correct results, especially when local and semilocal functionals are used. Despite intense theoretical focus over the years and given the inherent difficulties in developing universally accurate approximations to the unknown exchange-correlation functional for correlated systems, a systematically improvable framework purely within the context of DFT has not yet emerged\cite{Michael2017}. 
In contrast, many-body perturbation theories (MBPTs), as a promising alternative, provide a systematic framework to include electron correlations for ground-state as well as excited-state properties\cite{DFTvsMBPT_Angel2002}. 

Among different MBPTs, the random phase approximation (RPA)\cite{Exc_jellium_Langreth1997,RPA_Furche2001,RPA_Miyake2002,ACFDT_Furche2005,Fuchs_RPA_2005,ACFDT_hBN_Angel2006,ACFDT_RPA_Furche2008,RPA_review_Ren2012,SOSEX_Gruneis2009,RPA_variants_Angyan2011} is one of the simplest and most popular choices for calculating correlation energies beyond DFT. 
While many formulations and variants of RPA exist\cite{Exc_jellium_Langreth1997,RPA_Furche2001,RPA_Miyake2002,ACFDT_Furche2005,Fuchs_RPA_2005,ACFDT_hBN_Angel2006,ACFDT_RPA_Furche2008,RPA_review_Ren2012,SOSEX_Gruneis2009,RPA_variants_Angyan2011}, the framework based on the adiabatic connection fluctuation-dissipation theorem (ACFDT) is typically used in connection with advanced exchange-correlation functional for ground-state properties\cite{ACFDT_Niquet_2003,ACFDT_Furche2005,Fuchs_RPA_2005,ACFDT_RPA_Furche2008}. In addition, the RPA approach is connected to MBPT through the Klein functional\cite{Klein_1961}, evaluated at the level of the $GW$ approximation\cite{Leeuwen_PRA_2006}. 
The infinite sum of bubble diagrams in RPA provides the screening effects which are important for non-local correlation effects and van der Waals interactions. 
As a result, RPA is thus applicable to small-gap and metallic systems, unlike second-order Moller-Plesset perturbation theory (MP2) where the correlation energy diverges for systems with vanishing gaps\cite{Corr_energy_eGas_GellMann1957,MP2_solids_2010Gruneis}.

The conventional RPA energy for solids in a plane-wave basis requires a number of operations that scales quartically with the system size ($N$) and quadratically with the number of \textbf{k}-points ($N_{k}$), which makes its applications to large-scale systems rather expensive compared to DFT. 
Numerical techniques and optimizations have been introduced to reduce the formal scalings\cite{spacetime_GW_Rojas1995,spacetime_GW_enhancement_Steinbeck2000} and prefactors\cite{GW_virtuals_Bruneval2008,VASP_RPA_Kaltak2014,VASP_RPA_JCTC_Kaltak2014,sparse_sampling_Jia_2020,VASP_minimax_Kaltak2020,GW_virutals_Gao2016}. 
Particularly, the space-time approach\cite{spacetime_GW_Rojas1995} is proposed with a cubic scaling in terms of the system size and a linear scaling in terms of the number of \textbf{k}-points. 
This is achieved by transforming the computation of the polarizability on a real-space grid and the imaginary-time axis. 
Despite its appealing scaling, the space-time approach has only recently become competitive with other formulations as a result of new developments on efficient Fourier transforms on the imaginary axis\cite{VASP_RPA_Kaltak2014,VASP_RPA_JCTC_Kaltak2014,sparse_sampling_Jia_2020,VASP_minimax_Kaltak2020,DLR_Kaye2022}.  
Nevertheless, due to the large dimension of a real-space grid, the memory load is rather high, and the prefactors of the scaling laws are large compared to the quartic-scaling algorithm formulated in a canonical basis. 

A quartic-scaling algorithm for RPA can also be formulated in a localized single-particle basis with decomposition schemes of electron repulsion integrals (ERIs) such as Cholesky decomposition (CD)\cite{Cholesky_ERI_MOL_Beebe1977} and the resolution-of-the-indentity (RI) (also known as density fitting (DF)) technique\cite{DF_Werner2003,DF_Ren2012,GDF_MDF_Sun2017,RSDF_HongZhou2021}. 
Conceptually, both of these decomposition schemes factorize a rank-4 ERI tensor into a product of two rank-3 tensors by introducing an auxiliary basis whose size grows linearly with the system size. 
These types of decompositions result in a great amount of saving both in storage requirement and the number of operations, reducing the scaling of RPA in a localized basis from $O(N^{6})$ to $O(N^{4})$. 
One advantage of localized bases is their relatively compact size compared to plane-waves, so that the prefactors are significantly smaller compared to the space-time approach. 
For molecules and $\Gamma$-point supercells with small or intermediate sizes, the quartic-scaling algorithm in a localized basis could be more efficient compared to the cubic-scaling algorithm from the space-time approach, especially in the presence of core electrons. 
Further complexity reduction can be achieved by exploiting the sparsity of the fitting coefficients from DF with the overlap or the Coulomb-attenuated metric\cite{CP2K_cubicRPA_2016,lineaRPA_AO_Schurkus2016,linearRPA_Luenser2017}, and the locality of atomic orbitals\cite{linearDRPA_Kallay2015,lineaRPA_AO_Schurkus2016,linearRPA_Luenser2017}. 
Nevertheless, the assumptions of sparsity and locality of orbitals are valid only in the limit of large systems or for particular electronic properties which restrict their general applicability. 

In constrast, the $O(N^{4})$ algorithm for RPA based on DF/CD for ERIs is less appealing to solid-states systems due to the quadratic scaling with the number of \textbf{k}-points, which originates from the fact that the \textbf{k}-point indices in a rank-3 DF/CD tensor are not fully separable. 
Unlike the quartic scaling with the system size, the $O(N_{k}^{2})$ complexity can not be straightforwardly alleviated by exploiting sparsity or locality of orbitals. 
Furthermore, the lack of customized atomic orbitals for solids hinders convergence to the complete basis set limit. 
Standard Gaussian-type orbitals (GTOs) optimized in an atomic environment cannot be directly transferred to periodic systems due to the linear dependency problems in the presence of diffuse orbitals\cite{GTO_AFQMC_Miguel2020,GTO_Zhou2021,ccgto_Ye2022}. 
The problem becomes even more severe for DF whose accuracy relies on existence of a customized auxiliary basis set for a solid environment. 

An alternative decomposition of ERIs is tensor hypercontraction (THC) proposed by Hohenstein and co-workers\cite{THC_one_Martinez2012}. 
THC expresses an ERI tensor as a product of five matrices such that full separation of the four orbital indices in an ERI is obtained. 
There are different approaches to achieve the THC factorization such as the PARAFAC (PF) THC\cite{THC_one_Martinez2012}, least-squares (LS) THC\cite{LSTHC_Sherrill2012,THC_DVR_Sherrill2013,LSTHC_MP2_Schumacher2015}, and interpolative separable density fitting (ISDF)\cite{ISDF_Lu2015,ISDF_Bloch_Lu2016}. 
Due to the full separation of the four orbital indices, THC is able to further reduce the memory loads and the number of operations compared to DF and CD approaches. 
THC has been extensively applied to molecules and $\Gamma$-point supercells in the context of hybrid functionals\cite{ISDF_QRCP_hybrid_Hu2017,ISDF_CVT_hybrid_Lin2018,ISDF_hybridDFT_NAO_Qin2020,ISDF_CVT_hybrid_NAO_Qin2020}, Hartree-Fock (HF) theory\cite{rPS_HF_Sharma2022}, coupled-cluster (CC) theory\cite{THC_three_Martinez2012,THC_CC2_Hohenstein2013,THC_EOMCC2_Hohenstein2013,LSTHC_CCSD_Parrish2014}, MP2 and MP3\cite{THC_one_Martinez2012,LSTHC_Sherrill2012,THC_DVR_Sherrill2013,LSTHC_MP2_Schumacher2015,THC_SOS_MP2_one_Song2016,THC_SOS_MP2_two_Song2017,THC_SOS_MP2_gradient_Song2017,THC_MP3_Joonho2020}, RPA\cite{ISDF_RPA_model_Lu2017,Separable_RI_RPA_Blase2019}, $GW$\cite{TDDFT_G0W0_ISDF_Gao2020,G0W0_COHSEX_ISDF_Ma2021,separable_RI_G0W0_Blase2021}, and auxiliary-field quantum Monte-Carlo (AFQMC)\cite{AFQMC_ISDF_Miguel2019}. 
In contrast, for periodic calculations with \textbf{k}-point sampling, THC has only been used to accelerate the computation of hybrid functionals\cite{ISDF_hybridDFT_kpts_2022}. 

In this paper, we present an efficient algorithm for RPA with \textbf{k}-point sampling in the framework of THC. 
The formulation is based on a revised ISDF procedure for Bloch orbitals with a momentum-dependent THC auxiliary basis, resulting in full separation of both the orbital and the \textbf{k}-point indices. 
Both the preparation steps of ERIs and the evaluation steps of RPA energy can be performed at the cost of $O(N_{k}N^{3})$ in the number of operations and $O(N_{k}N^{2})$ in memory load without assumptions on sparsity or locality of orbitals. 
In the evaluation of RPA energy, the largest dimension of $N$ corresponds to the size of the THC auxiliary basis rather than the size of a real-space grid, which makes the prefactors much smaller compared to the standard space-time approach. 
We analyze the error convergence of ERIs and RPA energy with respect to the size of the THC auxiliary basis for different numbers of virtual orbitals, different numbers of \textbf{k}-points, and different sizes of unit cells. 

The paper is organized as follows. 
Sec.~\ref{sec:eri} introduces ERIs for periodic calculations, and Sec.~\ref{sec:thc} presents \textbf{k}-point THC via our revised ISDF procedure. 
In Sec.~\ref{sec:RPA}, we discusses the formulation of RPA in the framework of THC with \textbf{k}-point sampling. 
We then summarize the computational details in Sec.~\ref{sec:comp_detail}, and then reports results of our implementations of THC and RPA in Sec.~\ref{sec:results}. 
Lastly, our conclusion is presented in Sec.~\ref{sec:conclusion}. 

\section{Electron repulsion integrals \label{sec:eri}}
In the presence of translational symmetry, a suitable single-particle basis for the electronic Hamiltonian of a crystalline system is the Bloch orbital: 
\begin{align}
\phi^{\textbf{k}}_{i}(\textbf{r}) = u^{\textbf{k}}_{i}(\textbf{r})e^{i\textbf{kr}}
\label{eq:bloch_orbital}
\end{align}
where the superscripts $\{\textbf{k}\}$ denote crystal momenta, the subscripts are referred to as orbital indices, and $u^{\textbf{k}}_{i}(\textbf{r})$ are periodic functions with respect to lattice translations. 
In practice, the Bloch orbitals could be ``downfolded'' KS orbitals from a plane-wave basis, periodic Gaussian basis functions, or any other properly symmetry adapted set of basis functions. 
The ERIs in this basis are defined as 
\begin{align}
V^{\textbf{k}_{i}\textbf{k}_{j}\textbf{k}_{k}\textbf{k}_{l}}_{\ i\ j\ \ k\ l} = 
\int d\textbf{r} \int d\textbf{r}' \phi^{\textbf{k}_{i}*}_{i}(\textbf{r})\phi^{\textbf{k}_{j}}_{j}(\textbf{r})\frac{1}{|\textbf{r}-\textbf{r}'|}\phi^{\textbf{k}_{k}*}_{k}(\textbf{r}')\phi^{\textbf{k}_{l}}_{l}(\textbf{r}') 
\label{eq:eri}
\end{align}
where crystal momenta live in the first Brillouin zone with the assumption of momentum conservation, i.e. $\textbf{k}_{i} - \textbf{k}_{j} = \textbf{k}_{l} - \textbf{k}_{k} + \textbf{G}$ and $\textbf{G}$ is a reciprocal lattice vector. 

In first-principles calculations, the electronic Hamiltonian is constructed by discretizing the first Brillouin zone with a finite number of \textbf{k}-points ($N_{k}$) and truncating the Hilbert space using a fixed number of orbitals per unit cell ($N_{\mathrm{orb}}$). 
The size of the ERI tensor thus grows cubically with $N_{k}$ and quartically with $N_{\mathrm{orb}}$, which becomes a bottleneck both in computation cost and memory requirements as the system size increases. 
Furthermore, any operation on this bulky rank-4 tensor would lead to a poor scaling in terms of the number of operations due to the inseparability of the orbital and momentum indices. 

\section{Tensor hypercontraction\label{sec:thc}}
We assume the following tensor hypercontraction (THC) representation of Eq.~\ref{eq:eri} in a generic Bloch basis set: 
\begin{align}
V&^{\textbf{k}_{i}\textbf{k}_{j}\textbf{k}_{k}\textbf{k}_{l}}_{\ i\ j\ \ k\ l} \approx \sum_{\mu\nu}X^{\textbf{k}_{i}*}_{\mu i}X^{\textbf{k}_{j}}_{\mu j}V^{\textbf{q}}_{\mu\nu}X^{\textbf{k}_{k}*}_{\nu k}X^{\textbf{k}_{l}}_{\nu l} \label{eq:bloch_eri_thc}
\end{align}
where the momentum transferred between the Bloch pair densities is folded back to the first Brillouin zone ($\textbf{q} = \textbf{k}_{i} - \textbf{k}_{j} + \textbf{G} = \textbf{k}_{l} - \textbf{k}_{k} + \textbf{G}'$), and the greek letters denote the auxiliary basis introduced in the THC decomposition. 
For a given size of the auxiliary basis ($N_{\mu}$), the procedure of a THC decomposition consists of the determination of $\textbf{X}^{\textbf{k}}$ and $\textbf{V}^{\textbf{q}}$ matrices. 
When $N_{\mu}$ is smaller than $N_{\mathrm{orb}}^{2}$, Eq.~\ref{eq:bloch_eri_thc} corresponds to a low-rank approximation to an ERI tensor. 
In practice, $N_{\mu} = O(N_{\mathrm{orb}})$ is expected to achieve good accuracy due to the low-rank structure of the ERI tensor. 
The expression of Eq.~\ref{eq:bloch_eri_thc} provides full separation of the orbital and momentum indices which not only reduces the memory requirements but also enables a low-scaling algorithm for RPA energy (see Sec.~\ref{sec:RPA}). 
In this work, we proposed a revised ISDF procedure, based on the works from Lu and coworkers\cite{ISDF_Lu2015,ISDF_Bloch_Lu2016}, to construct the THC factorization with a momentum-dependent auxiliary basis. 

\subsection{Interpolative separable density fitting for solids\label{subsec:isdf}}
For a fixed transferred momentum $\textbf{q}$ that lives in the first Brillouin zone, we view the Bloch pair densities for arbitrary $\textbf{k}$-points as a matrix $\rho^{\textbf{q}}(\textbf{k}ij, \textbf{r}) = \phi^{\textbf{k}-\textbf{q}*}_{i}(\textbf{r})\phi^{\textbf{k}}_{j}(\textbf{r})$, and then we perform an interpolative decomposition (ID)\cite{ID_Cheng2005,ID_Liberty2007} to $\bm{\rho}^{\textbf{q}}$: 
\begin{align}
\rho^{\textbf{q}}(\textbf{k}ij, \textbf{r}) = \phi^{\textbf{k}-\textbf{q}*}_{i}(\textbf{r})\phi^{\textbf{k}}_{j}(\textbf{r}) \approx \sum_{\mu}\phi^{\textbf{k}-\textbf{q}*}_{i}(\textbf{r}_{\mu})\phi^{\textbf{k}}_{j}(\textbf{r}_{\mu})\zeta^{\textbf{q}}(\mu, \textbf{r})
\label{eq:id_rho}
\end{align}
where $\{\textbf{r}_{\mu}\}$ is a set of interpolating points, and $\{\zeta^{\textbf{q}}_{\mu}(\textbf{r})\}$ are interpolating vectors that interpolate pair densities to an arbitrary real-space point $\textbf{r}$ from $\{\textbf{r}_{\mu}\}$. 
The number of interpolating points ($N_{\mu}$) can either be an input parameter or determined on-the-fly for given accuracy. 
Since the size of the real-space grid ($N_{r}$) scales linearly with the number of electrons ($N_{e}$), $N_{\mu}$ is expected to grow as $\mathcal{O}(N_{e})$ as well. 
Due to the periodicity of the pair densities in the momentum space, it is easy to verify that $\zeta^{\textbf{q}}_{\mu}(\textbf{r})$ is also a Bloch function, i.e. $\zeta^{\textbf{q}+\textbf{G}}_{\mu}(\textbf{r}) = \zeta^{\textbf{q}}_{\mu}(\textbf{r})$. 
The structure of Eq.~\ref{eq:id_rho} resembles the widely-used density fitting decomposition\cite{DF_Werner2003,DF_Ren2012,GDF_MDF_Sun2017,RSDF_HongZhou2021} if one identifies the interpolating vectors $\{\zeta^{\textbf{q}}_{\mu}(\textbf{r})\}$ as the auxiliary basis set. 
However, the fitted coefficients are now separable both in the orbital and the $\textbf{k}$-point indices, and the auxiliary basis set is numerically determined during the fitting procedure rather than taken from a set of predefined functions. 

This fitting procedure is performed independently for each \textbf{q}-point to generate a set of $\textbf{q}$-dependent interpolating basis $\{\zeta^{\textbf{q}}_{\mu}(\textbf{r})\}$. 
In principle, the optimal interpolating points should also be $\textbf{q}$-dependent. 
However, as shown in the Supporting Information, we empirically found that taking $\{\textbf{r}_{\mu}\}$ from $\textbf{q=0}$ consistently results in comparable accuracy as in the case that uses \textbf{q}-dependent interpolating points. 

Finally, a THC representation of ERIs is obtained by inserting Eq.~\ref{eq:id_rho} into Eq.~\ref{eq:eri}: 
\begin{subequations}
\begin{align}
V^{\textbf{k}_{i}\textbf{k}_{j}\textbf{k}_{k}\textbf{k}_{l}}_{\ i\ j\ \ k\ l} &\approx \sum_{\mu\nu}\phi^{\textbf{k}_{i}*}_{i}(\textbf{r}_{\mu})\phi^{\textbf{k}_{j}}_{j}(\textbf{r}_{\mu}) \Big[\int d\textbf{r}  \int d\textbf{r}' \zeta^{-\textbf{q}}_{\mu}(\textbf{r})\frac{1}{|\textbf{r}-\textbf{r}'|}\zeta^{\textbf{q}}_{\nu}(\textbf{r}') \Big] \phi^{\textbf{k}_{k}*}_{k}(\textbf{r}_{\nu})\phi^{\textbf{k}_{l}}_{l}(\textbf{r}_{\nu})\label{eq:eri_isdf_a}\\
&=\sum_{\mu\nu}X^{\textbf{k}_{i}*}_{\mu i}X^{\textbf{k}_{j}}_{\mu j}V^{\textbf{q}}_{\mu\nu}X^{\textbf{k}_{k}*}_{\nu k}X^{\textbf{k}_{l}}_{\nu l}\label{eq:eri_isdf_b}
\end{align}
\label{eq:eri_isdf}
\end{subequations}
where we define $\textbf{q} = \textbf{k}_{i} - \textbf{k}_{j} + \textbf{G} = \textbf{k}_{l} - \textbf{k}_{k} + \textbf{G}'$, and 
\begin{subequations}
\begin{align}
&X^{\textbf{k}_{i}}_{\mu i} = \phi^{\textbf{k}_{i}}_{i}(\textbf{r}_{\mu}), \label{eq:X_thc} \\ 
&V^{\textbf{q}}_{\mu\nu} = \int d\textbf{r} \int d\textbf{r}' \zeta^{-\textbf{q}}_{\mu}(\textbf{r})\frac{1}{|\textbf{r}-\textbf{r}'|}\zeta^{\textbf{q}}_{\nu}(\textbf{r}').  \label{eq:V_thc}
\end{align}
\end{subequations}
The accuracy of Eq.~\ref{eq:eri_isdf_b} is controlled by the accuracy of ISDF procedure (Eq.~\ref{eq:id_rho}) which can be systematically improved by increasing $N_{\mu}$.  

What remains is how to obtain the ID representation in Eq.~\ref{eq:id_rho}. 
The standard procedure of ID consists of first selecting the interpolating points and then solving a least-squares problem to obtain the interpolating vectors\cite{ID_Cheng2005}. 
In our implementation, we select the interpolating points using the recently-proposed scheme based on the Cholesky decomposition of the THC metric matrix at \textbf{q=0}\cite{LSTHC_Matthews2020} (see Sec.~\ref{subsubsec:chol}). 
Once the interpolating points are chosen, the interpolating vectors are obtained from the least-squares solution of the following over-determined set of linear equations: 
\begin{align}
\textbf{C}^{\textbf{q}}\bm{\Theta}^{\textbf{q}} = \textbf{Z}^{\textbf{q}}
\label{eq:thc_ls}
\end{align}
where 
\begin{align}
Z^{\textbf{q}}_{\nu\textbf{r}} &= \sum_{\textbf{k}ij}\rho^{\textbf{q}*}(\textbf{k}ij,\textbf{r}_{\nu})\rho^{\textbf{q}}(\textbf{k}ij, \textbf{r}), \label{eq:Z} \\
C^{\textbf{q}}_{\nu\mu} &= \sum_{\textbf{k}ij}\rho^{\textbf{q}*}(\textbf{k}ij,\textbf{r}_{\nu})\rho^{\textbf{q}}(\textbf{k}ij, \textbf{r}_{\mu}), \label{eq:C} \\ 
\Theta^{\textbf{q}}_{\mu\textbf{r}} &= \zeta^{\textbf{q}}_{\mu}(\textbf{r}). 
\end{align}

Due to the separability of the orbital indices in $\bm{\rho}^{\textbf{q}}$, the evaluation of Eq.~\ref{eq:Z} scales as $\mathcal{O}(N_{k}N_{\mathrm{orb}}N_{\mu}N_{r} + N_{k}\ln{N_{k}}N_{\mu}N_{r})$. 
Once $\textbf{Z}^{\textbf{q}}$ and $\textbf{C}^{\textbf{q}}$ are assembled, solving the linear system scales as $\mathcal{O}(N_{k}N_{\mu}^{2}N_{r} + N_{k}N_{\mu}^{3})$. 
Overall, the evaluation of interpolating vectors scales linearly with $N_{k}$ and cubically with the system size. 

\subsubsection{Cholesky-based approach for interpolating points\label{subsubsec:chol}}
In the Cholesky-based approach\cite{LSTHC_Matthews2020}, the interpolating points are selected through the pivoted Cholesky decomposition on the matrix $\textbf{S}^{\textbf{q}} = (\bm{\rho}^{\textbf{q}})^{\dag}\bm{\rho}^{\textbf{q}}$: 
\begin{align}
\textbf{S}^{\textbf{q}} = \bm{\Pi}^{\textbf{q}}(\textbf{R}^{\textbf{q}})^{\dag}\textbf{R}^{\textbf{q}}(\bm{\Pi}^{\textbf{q}})^{-1}
\label{eq:pivoted_chol}
\end{align}
where $\bm{\Pi}^{\textbf{q}}$ is the pivoting matrix and $\textbf{R}^{\textbf{q}}$ consists of the Cholesky vectors with the diagonal elements in the descending order. 
For a given $N_{\mu}$, the interpolating points are then chosen to be those rows that correspond to the first $N_{\mu}$ pivots in $\bm{\Pi}^{\textbf{q}}$. 

This approach is a reformulation of QR factorization with column pivoting (QRCP) on the matrix $\bm{\rho}^{\textbf{q}}$, 
\begin{align}
\bm{\rho}^{\textbf{q}}\bm{\Pi}^{\textbf{q}} = \textbf{Q}^{\textbf{q}}\textbf{R}^{\textbf{q}},
\label{eq:QRCP}
\end{align}
which is the standard approach to select interpolating points in IDs\cite{ID_Cheng2005}. 
However, Eq.~\ref{eq:pivoted_chol} has several advantages over Eq.~\ref{eq:QRCP} from a numerical point of view. 
First of all, due to the separability of the orbital indices in $\bm{\rho}^{\textbf{q}}$, the evaluation of $\textbf{S}^{\textbf{q}}$ scales as $\mathcal{O}(N_{k}N_{\mathrm{orb}}N_{r}^{2})$ which is asymptotically cheaper than the evaluation of $\bm{\rho}^{\textbf{q}}$ ($\mathcal{O}(N_{\textbf{k}}^{2}N_{\mathrm{orb}}^{2}N_{r})$). 
Secondly, an direct QRCP on the matrix $\bm{\rho}^{\textbf{q}}$ is prohibitively expensive. 
Instead, the randomized algorithm of QRCP is typically implemented to reduce the cost to $\mathcal{O}(N_{k}N_{\mathrm{orb}}^{2}N_{r})$. 
On the other hand, the iterative procedure of pivoted Cholesky allows one to construct the matrix $\textbf{R}^{\textbf{q}}$ incrementally in a deterministic manner and terminate the algorithm once the error is below a user-defined threshold or the number of Cholesky vectors exceeds $N_{\mu}$. 
Therefore, the pivoted Cholesky decomposition on $\textbf{S}^{\textbf{q}}$ can be done at the cost of $\mathcal{O}(N_{k}N_{\mu}^{2}N_{r})$. 

\section{RPA energy \label{sec:RPA}}
The grand potential $\Omega$ of an interacting many-electron system can be expressed using the Klein functional\cite{Klein_1961} 
\begin{align}
\Omega_{\mathrm{K}}[G] = \Phi[G] + E_{\mathrm{H}} + \mathrm{Tr}[1 - G^{-1}_{0}G] - \mathrm{Tr}[\ln(-G^{-1})]
\label{eq:Klein_GP}
\end{align}
with the Hartree (Coulomb) energy $E_{\mathrm{H}}$, the non-interacting Green's function $G_{0}$, the interacting Green's function $G$, and the Luttinger-Ward functional $\Phi[G]$\cite{LW_1960}. 
The interacting Green's function relates to its non-interacting counterpart through the Dyson equation 
\begin{align}
G^{-1}(\omega) = G^{-1}_{0}(\omega) - \Sigma(\omega)
\label{eq:dyson}
\end{align}
in which $G$ and the self-energy $\Sigma$ are solved in a self-consistent manner. 

Since a self-consistent solution of the Dyson equation is computationally demanding, Eq.~\ref{eq:Klein_GP} is usually evaluated at an effective non-interacting Green's function such as the Kohn-Sham (KS) Green's function 
\begin{align}
G_{\mathrm{KS}}(\textbf{r},\textbf{r}', \omega) = \sum_{i}\frac{\psi^{*}_{i}(\textbf{r})\psi_{i}(\textbf{r}')}{\omega - \epsilon_{i} + i\delta}
\label{eq:G_KS}
\end{align}
where $\{\psi_{i}\}$ are the KS orbitals and $\{\epsilon_{i}\}$ are the KS single-particle energies. 
Inserting Eq.~\ref{eq:G_KS} into Eq.~\ref{eq:Klein_GP}, the single-particle nature of $G_{\mathrm{KS}}$ allows us to write down the relation\cite{Leeuwen_PRA_2006} 
\begin{align}
F[G_{\mathrm{KS}}] = \Omega_{K}[G_{\mathrm{KS}}] + \mu N = E_{\mathrm{HF}}[\{\psi_{i}\}] + \Phi_{c}[G_{\mathrm{KS}}] 
\label{eq:free_energy}
\end{align}
where $F$ is the Helmholtz free energy, $E_{\mathrm{HF}}[\{\psi_{i}\}]$ is the Hartree-Fock (HF) energy evaluated using the KS orbitals, and $\Phi_{c}[G_{\mathrm{KS}}]$ is the correlation part of the Luttinger-Ward function evaluated at $G_{\mathrm{KS}}$. 
In the RPA approximation, $\Phi_{c}$ is represented as a sum of bubble diagrams, 
\begin{subequations}
\begin{align}
\Phi^{\mathrm{RPA}}_{c}  &= -\frac{1}{2}\mathrm{Tr}\{ [(\chi_{0}V) + \frac{1}{2}(\chi_{0}V)^{2} + \frac{1}{3}(\chi_{0}V)^{3} + \frac{1}{4}(\chi_{0}V)^{4} + \dots] - (\chi_{0}V)\}  \label{eq:Phi_RPAa}\\
&= \frac{1}{2}\mathrm{Tr}\{ \ln[1 - \chi_{0}V] + \chi_{0}V \}, \label{eq:Phi_RPAb}
\end{align}
\label{eq:Phi_RPA}
\end{subequations}
where $\chi_{0} = G_{\mathrm{KS}}G_{\mathrm{KS}}$ is the KS polarizability, $V$ is the bare Coulomb interaction, and the $\mathrm{Tr}\{\}$ operator denotes a sum over all degrees of freedom. 
At the zero-temperature limit, Eq.~\ref{eq:Phi_RPA} is identical to the RPA correlation energy in the framework of adiabatic-connection fluctuation-dissipation theorem (ACFDT)\cite{ACFDT_Niquet_2003,ACFDT_Furche2005,Fuchs_RPA_2005,ACFDT_RPA_Furche2008}. 

\subsection{THC-HF \label{subsec:thc-hf}}
The HF energy expressed in a canonical basis $\{\psi_{i}\}$ is 
\begin{align}
E_{\mathrm{HF}}[\{\psi_{i}\}] = \frac{1}{2N_{k}}\sum_{\textbf{k}}\sum_{ij} \rho^{\textbf{k}}_{ij}(V^{\mathrm{HF}})^{\textbf{k}}_{ji}
\label{eq:hf_energy}
\end{align}
where $\bm{\rho}^{\textbf{k}}$ is the single-particle density matrix and $(\textbf{V}^{\mathrm{HF}})^{\textbf{k}}$ is the canonical HF potential. 
With the THC representation of ERIs from Eq.~\ref{eq:eri_isdf}, $(V^{\mathrm{HF}})^{\textbf{k}}_{ij} = J^{\textbf{k}}_{ij} + K^{\textbf{k}}_{ij}$ can be reformulated as 
\begin{subequations}
\begin{align}
J^{\textbf{k}}_{ij} &= \frac{2}{N_{k}}\sum_{\textbf{k}'}\sum_{ab}\rho^{\textbf{k}'}_{ab}V^{\textbf{k}\textbf{k}\textbf{k}'\textbf{k}'}_{ijba} \label{eq:thc_Ja}\\
&= \frac{2}{N_{k}}\sum_{\textbf{k}'}\sum_{\mu\nu}X^{\textbf{k}*}_{i\mu}X^{\textbf{k}}_{j\mu}V^{\textbf{q=0}}_{\mu\nu}\sum_{ab}X^{\textbf{k}'}_{a\nu}\rho^{\textbf{k}'}_{ab}X^{\textbf{k}'*}_{b\nu} \label{eq:thc_Jb}\\ 
& = \sum_{\mu}X^{\textbf{k}*}_{i\mu}\big\{ \frac{2}{N_{k}}\sum_{\textbf{k}'} \sum_{\nu}\rho^{\textbf{k}'}(\textbf{r}_{\nu}, \textbf{r}_{\nu})V^{\textbf{q=0}}_{\mu\nu}\big\}X^{\textbf{k}}_{j\mu} \label{eq:thc_Jc}
\end{align}
\label{eq:thc_J}
\end{subequations}
and 
\begin{subequations}
\begin{align}
K^{\textbf{k}}_{ij} &= -\frac{1}{N_{k}}\sum_{\textbf{q}}\sum_{ab}\rho^{\textbf{k-q}}_{ab}V^{\textbf{k},\textbf{k-q},\textbf{k-q},\textbf{k}}_{i, a, b, j} \label{eq:thc_Ka}\\
&= \frac{-1}{N_{k}}\sum_{\textbf{q}}\sum_{\mu\nu}X^{\textbf{k}*}_{i\mu}V^{\textbf{q}}_{\mu\nu}X^{\textbf{k}}_{j\nu}\sum_{ab}X^{\textbf{k-q}}_{a\mu}\rho^{\textbf{k-q}}_{ab}X^{\textbf{k-q}*}_{b\nu} \label{eq:thc_Kb}\\
&= \sum_{\mu\nu}X^{\textbf{k}*}_{i\mu} \big\{ \frac{-1}{N_{k}}\sum_{\textbf{q}}\rho^{\textbf{k-q}}(\textbf{r}_{\mu}, \textbf{r}_{\nu})V^{\textbf{q}}_{\mu\nu}\big\} X^{\textbf{k}}_{j\nu} \label{eq:thc_Kc}
\end{align}
\label{eq:thc_K}
\end{subequations}
in which $\textbf{J}^{\textbf{k}}$ is the Coulomb term, $\textbf{K}^{\textbf{k}}$ is the exchange term, $\bm{\rho}^{\textbf{k}}$ is the single-particle density matrix, and $\rho^{\textbf{k}}(\textbf{r}_{\mu}, \textbf{r}_{\nu})$ is referred to as the electron density evaluated on the THC interpolating points: 
\begin{align}
\rho^{\textbf{k}}(\textbf{r}_{\mu}, \textbf{r}_{\nu})  = \sum_{ab}X^{\textbf{k}}_{a\mu}\rho^{\textbf{k}}_{ab}X^{\textbf{k}*}_{b\nu} = \sum_{ab}\phi^{\textbf{k}}_{a}(\textbf{r}_{\mu})\rho^{\textbf{k}}_{ab}\phi^{\textbf{k}*}_{b}(\textbf{r}_{\nu}). 
\label{eq:rho_thc}
\end{align}

The most time-consuming part in THC-HF is Eq.~\ref{eq:thc_Kc} which scales as $O(N_{k}\ln{N_{k}}N_{\mu}^{2}+N_{k}N_{\mathrm{orb}}N_{\mu}^{2})$. Here, the logarithmic complexity comes from the fast Fourier transform (FFT) convolution in the momentum space. 
Therefore, the evaluation of THC-HF scales linearly with $N_{k}$ and cubically with the system size. 
This complexity is asymptotically much better than other approaches formulated in a canonical basis, such as those based on the Gaussian density-fitting technique\cite{DF_Ren2012} or the Cholesky decomposition\cite{Cholesky_ERI_MOL_Beebe1977}. 
In addition, compared to the real-space formalism which has the same formal scaling, the prefactor of THC-HF is several orders of magnitude smaller since $N_{\mu} \ll N_{r}$. 
Even though the preparation steps for obtaining the THC decomposition of ERIs still acquires a large prefactor from the dimension of the real-space grid, this step is only done at once in the beginning of the calculation, no matter the number of self-consistent cycles in THC-HF. 

\subsection{THC-RPA \label{subsec:thc-rpa}}
Similar to HF, the non-interacting polarizability can be reformulated using the THC interpolating points and the THC auxiliary basis on the imaginary-time axis. 
On a real-space grid, the polarizability reads  
\begin{align}
\chi_{0}(\textbf{r}, \textbf{r}'; \tau) &= G(\textbf{r}, \textbf{r}'; \tau)G(\textbf{r}', \textbf{r}; -\tau) \nonumber\\
&= \sum_{\textbf{k}\textbf{q}}\sum_{abcd} \phi^{\textbf{k}}_{a}(\textbf{r})\phi^{\textbf{k}-\textbf{q}*}_{c}(\textbf{r})G^{\textbf{k}}_{ab}(\tau)G^{\textbf{k}-\textbf{q}}_{dc}(-\tau)\phi^{\textbf{k}-\textbf{q}}_{d}(\textbf{r}')\phi^{\textbf{k}*}_{b}(\textbf{r}') \nonumber\\ 
&= \sum_{\textbf{q}\textbf{k}}\sum_{\mu\nu}\zeta^{\textbf{q}}_{\mu}(\textbf{r})G^{\textbf{k}}(\textbf{r}_{\mu}, \textbf{r}_{\nu}; \tau)G^{\textbf{k}-\textbf{q}}(\textbf{r}_{\nu}, \textbf{r}_{\mu}; -\tau)\zeta^{\textbf{q}*}_{\nu}(\textbf{r}') \nonumber\\
&= \sum_{\textbf{q}}\sum_{\mu\nu}\zeta^{\textbf{q}}_{\mu}(\textbf{r})\chi_{0}^{\textbf{q}}(\textbf{r}_{\mu}, \textbf{r}_{\nu}; \tau)\zeta^{\textbf{q}*}_{\nu}(\textbf{r}')
\label{eq:chi_rr'}
\end{align}
where 
\begin{align}
G^{\textbf{k}}(\textbf{r}_{\mu}, \textbf{r}_{\nu}; \tau) &= \sum_{ab}\phi^{\textbf{k}}_{a}(\textbf{r}_{\mu})G^{\textbf{k}}_{ab}(\tau)\phi^{\textbf{k}*}_{b}(\textbf{r}_{\nu}), \label{eq:G_thc}\\
\chi^{\textbf{q}}_{0}(\textbf{r}_{\mu}, \textbf{r}_{\nu}; \tau) &= \sum_{\textbf{k}}G^{\textbf{k}}(\textbf{r}_{\mu}, \textbf{r}_{\nu}; \tau)G^{\textbf{k-q}}(\textbf{r}_{\nu}, \textbf{r}_{\mu}; -\tau). 
\label{eq:chi_thc}
\end{align}
Inserting Eq.~\ref{eq:chi_rr'} into Eq.~\ref{eq:Phi_RPAa}, we recast the first-order term into 
\begin{subequations}
\begin{align}
-\frac{1}{2}\mathrm{Tr}&\{\chi_{0}V\} = -\frac{1}{2\beta}\sum_{n}\int d\textbf{r}\int d\textbf{r}' \chi_{0}(\textbf{r}, \textbf{r}'; i\Omega_{n})V(\textbf{r}, \textbf{r}') \label{eq:RPA_1sta}\\
&=-\frac{1}{2\beta}\sum_{n}\sum_{\textbf{q}}\sum_{\mu\nu}\chi_{0}^{\textbf{q}}(\textbf{r}_{\mu}, \textbf{r}_{\nu}; i\Omega_{n}) V^{\textbf{q}}_{\nu\mu} \label{eq:RPA_1stb}
\end{align}
\label{eq:RPA_1st}
\end{subequations}
in which $V^{\textbf{q}}_{\nu\mu}$ is defined in Eq.~\ref{eq:V_thc}. 
Similar reformulation can be applied to the higher-order terms of Eq.~\ref{eq:Phi_RPAa}, and the final expression of $\Phi^{\mathrm{RPA}}_{c}$ reads 
\begin{align}
\Phi^{\mathrm{RPA}}_{c} = \frac{1}{2\beta}\sum_{n}\sum_{\textbf{q}}\sum_{\mu}\{ \ln[1 - \chi_{0}^{\textbf{q}}(i\Omega_{n})V^{\textbf{q}}] + \chi_{0}^{\textbf{q}}(i\Omega_{n})V^{\textbf{q}} \}_{\mu\mu}.
\label{eq:phi_thc_rpa}
\end{align}

The formal scaling of Eqs.~\ref{eq:G_thc} and~\ref{eq:phi_thc_rpa} scales as $O(N_{\tau}N_{k}N_{\mathrm{orb}}N_{\mu}^{2})$ and $O(N_{\Omega}N_{k}N^{3}_{\mu})$ respectively, and Eq.~\ref{eq:chi_thc} can be evaluated using the FFT convolution at the cost of $O(N_{\tau}N_{k}\ln N_{k}N_{\mu}^{2})$. 
Therefore, each step formally scales linearly with $N_{k}$ and cubically with the system size. 
Particularly, Eq.~\ref{eq:phi_thc_rpa} would be the most time-consuming step, assuming the sizes of the Matsubara frequencies and the imaginary-time mesh are similar. 
Note that the low-scaling algorithm of THC-RPA is a consequence of the full separability in the orbital and $\textbf{k}$-point indices from the THC factorization of ERIs. 
This formalism does not rely on any assumption on sparsity, and it can be applied to any generic Bloch orbitals as long as there is an reasonably compact ID for the pair densities. 

In addition to the formal cubic scaling, the THC-RPA algorithm has a much smaller prefactor compared to the space-time formalism. 
This can be seen from the construction of the non-interacting polarizability (Eqs.~\ref{eq:G_thc} and~\ref{eq:chi_thc}) in which the two formalisms look almost the same except that the real-space grid in the space-time formalism is replaced by the THC interpolating points. 
Since the dimension of the later is often orders of magnitude smaller than the former, THC-RPA gains further speedup even compared to the space-time formalism. 

\section{Computational Details\label{sec:comp_detail}}
For all the data presented in this work, we choose the KS orbitals from a DFT calculation as the single-particle Bloch basis. 
Unless mentioned otherwise, the total number of KS states is taken to be 8 times of the number of electrons per unit cell. 
All functions in this basis set are used to construct the electronic Hamiltonian in THC factorization and compute the HF and the RPA correlation energy. 

All DFT calculations are performed with the Perdew-Burke-Ernzerhof (PBE) exchange-correlation functional\cite{PBE_Perdew1996} using \texttt{Quantum Espresso}\cite{QE_Giannozzi2009,QE_Giannozzi2017,QE_Giannozzi2020}. 
Core electrons are described by norm-conserving pseudopotentials optimized for the PBE functional\cite{ONCVPP_Hamann2013,SG15ONCV_Schlipf2015,pseudodojo_2018}, and the kinetic energy cutoff is set to 55 a.u. for all systems unless mentioned otherwise. 

RPA calculations are performed exclusively on the imaginary axes at inverse temperature $\beta = 2000$ a.u. ($T \approx 158$ K). 
Dynamic quantities, including fermionic and bosonic functions, are expanded into the intermediate representation (IR)\cite{IR_Hiroshi_2017} with sparse sampling on both the imaginary-time and Matsubara frequency axes\cite{sparse_sampling_Jia_2020}. 
Both the IR basis and the sampling points are generated using \texttt{sparse-ir}\cite{spare_ir_Markus2023} open-source software package. 

\section{Results \label{sec:results}}
In this section, we present the results of our implementation of the THC decomposition of ERIs, THC-HF and THC-RPA. 
To facilitate the comparison between different physical systems and basis sets, we define the metrics $\alpha = N_{\mathrm{\mu}}/N_{\mathrm{orb}}$ which represents the size of the auxiliary basis as a multiple of the size of the single-particle Bloch orbitals. 

\subsection{ERI comparison between different factorization schemes \label{subsec:thc_eri}}
\begin{figure}[tbh!]
\begin{center}
\includegraphics[width=0.35\textwidth]{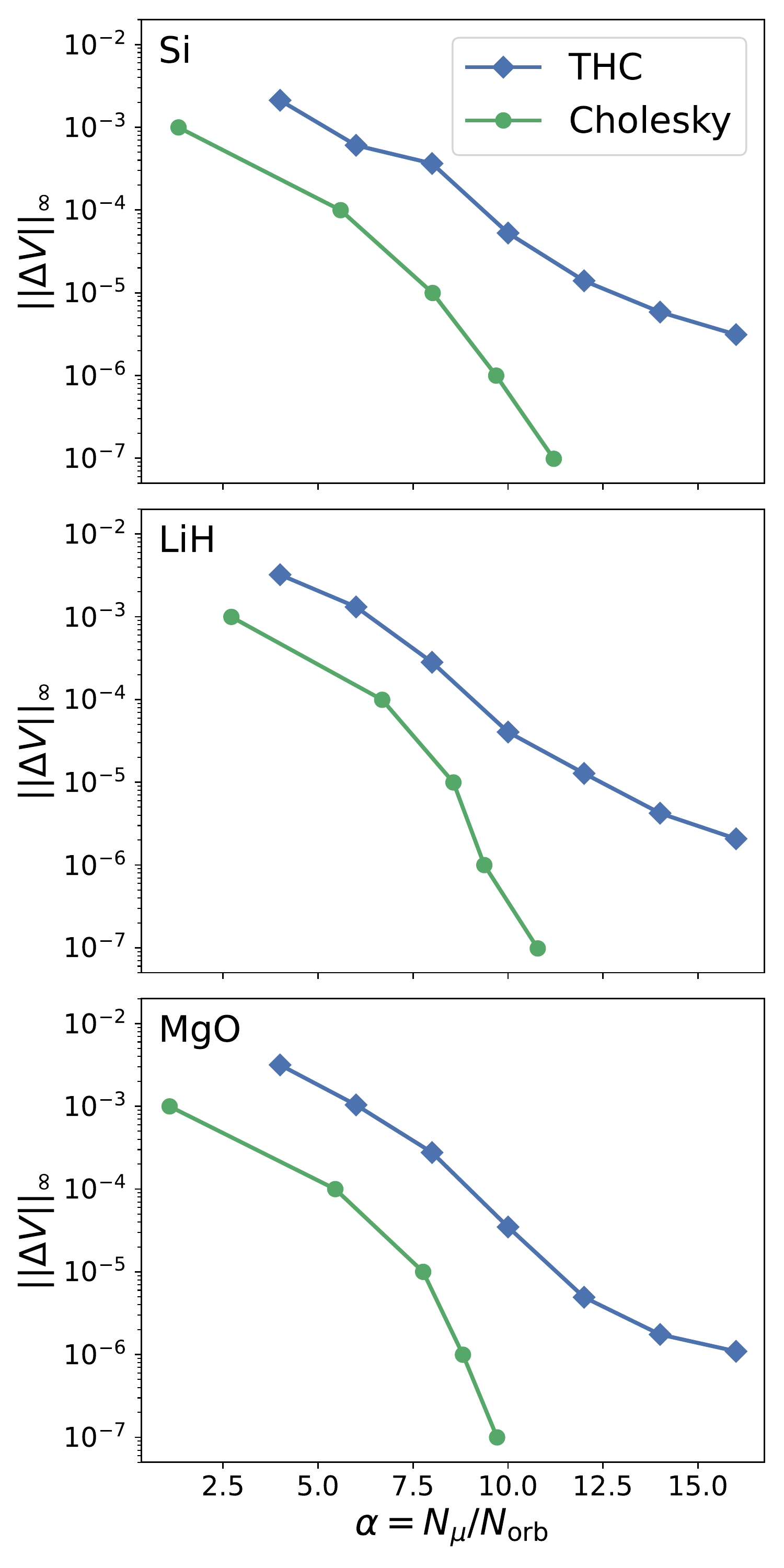}
\caption{Maximum error (a.u.) of ERIs for all orbital blocks calculated using different decomposition methods with respect to the sizes of the auxiliary bases ($\alpha  = N_{\mathrm{\mu}}/N_{\mathrm{orb}}$) for Si (top), LiH (middle), and MgO (bottom). }
\label{fig:eri_conv}
\end{center}
\end{figure}

We first investigate the error of the THC representation of ERIs for a given size of the auxiliary basis. 
Fig.~\ref{fig:eri_conv} shows the maximum error of the ERI tensor $V^{\textbf{k}_{i}\textbf{k}_{j}\textbf{k}_{k}\textbf{k}_{l}}_{\ i\ j\ \ k\ l}$, including the occupied-occupied, the occupied-virtual, and the virtual-virtual orbital blocks, for selected physical systems calculated from the THC, and the Cholesky decomposition at different sizes of the auxiliary bases. 
In the following, the auxiliary basis for Cholesky decomposition is referred to as the Cholesky vectors.  
The selected systems are chosen to be Si, LiH, and MgO with increasing band gaps on a $2\times2\times2$ $\Gamma$-centered Monkhorst-Pack grid. 
Such a small $\textbf{k}$-grid is to alleviate the high computation cost of assembling the full ERI tensor from the decomposed forms. 
As will be demonstrated in the next section (Fig.~\ref{fig:rpa_kpts_scell_conv}), the convergence of THC should remain consistent no matter the size of the $\textbf{k}$-mesh. 
The reference data are calculated from Cholesky decomposition with convergence tolerance equals to $10^{-8}$. 

Overall, the two decomposition schemes show monotonic convergence as the sizes of the auxiliary bases increase. 
The similar convergence behavior among the three selected systems with different numbers of orbitals suggests that the rank of the full ERI tensor only grows linearly with the system size, independent to the details of the system, \textit{e}.\textit{g}. the size of the band gaps. 
Among the two factorization schemes, Choleksy decomposition consistently shows faster convergence since it does not require a fully separable form in the orbital and $\textbf{k}$-point indices. For THC, we found that $\alpha_{\mathrm{THC}}=8$ already gives us accuracy better than 1 mHartree for all orbital blocks in ERIs. 
The consistent accuracy for different orbital blocks is because both of the decomposition schemes treat the occupied and the virtual orbitals on an equal footing. 
Therefore, both of the Cholesky and the THC decomposition are applicable to not only mean-field calculations but also correlated methods which involve the occupied-virtual and the virtual-virtual interactions. 

Despite a larger auxiliary basis, THC is still computationally favorable compared to Cholesky decomposition due to the linear scaling with the number of $\textbf{k}$-points and the cubic scaling with the system sizes. 
From the perspective of memory usage, the fully separable form of THC reduces the storage requirement from $O(N_{k}^{2}N_{\mathrm{orb}}^{2}N_{\mu})$ to $O(N_{k}N_{\mu}^{2})$ compared to Cholesky decomposition. 
Such memory reduction allows for the possibility to compute and store the full decomposed ERI on-the-fly and avoid I/O entirely. 

\subsection{RPA free energy}
\begin{figure}[tbh!]
\begin{center}
\includegraphics[width=0.58\textwidth]{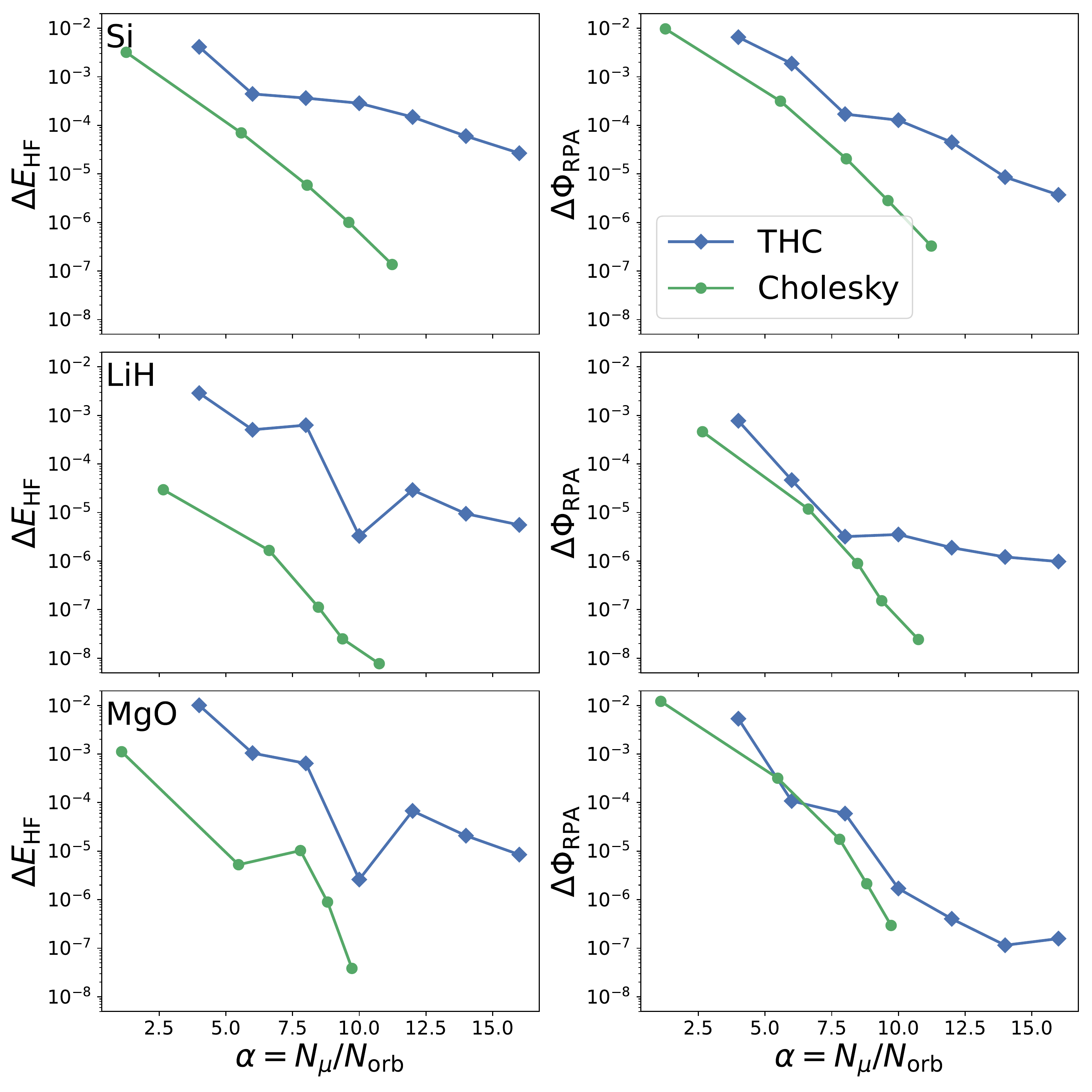} 
\caption{Convergence of HF (left column) and RPA correlation energy per atom (right column) for Si (top), LiH (middle), MgO (bottom) with respect to the sizes of the auxiliary bases ($\alpha =  N_{\mathrm{\mu}}/N_{\mathrm{orb}}$) from different decomposition schemes of ERIs. The unit of energy is Hartree. }
\label{fig:rpa_conv}
\end{center}
\end{figure}

Next, we analyze the convergence of the RPA free energy with respect to the size of the auxiliary basis. 
Fig.~\ref{fig:rpa_conv} shows the error of the HF (Eq.~\ref{eq:hf_energy}) and the RPA correlation energy (Eq.~\ref{eq:phi_thc_rpa}) per atom, calculated using ERIs in the THC decomposition as a function of $\alpha = N_{\mu}/N_{\mathrm{orb}}$. 
For consistency, we choose the same physical systems on the same $\textbf{k}$-mesh as in Sec.~\ref{subsec:thc_eri}. 
We also show the HF and the RPA results from the Cholesky decomposed ERI, denoted as Chol-HF and Chol-RPA, using our in-house library for many-body theory which closely follows the finite-temperature implementation in Ref.\citenum{scGW_CNY2022}. 

Both of our implementations are able to systematically converge to the same results within given accuracy as $\alpha$ increases since ISDF and Cholesky decomposition are both systematically controlled approximations. 
Such a high accuracy calculation is not possible with the conventional density fitting techniques in which the error is subject to the choice of a pre-defined auxiliary basis set. 
Compared to the error in ERIs, the convergence of energetics is less smooth since the errors coming from the ERI factorization propagate non-linearly in the energy evaluation. 
However, the overall trend remains the same, i.e. one can achieve approximately 1 mHartree and 0.01 mHartree accuracy at $\alpha_{\mathrm{THC}} = 8$ and $16$, respectively. 
Unlike HF, the RPA correlation energy requires the information of interactions from virtual orbitals. 
The consistent accuracy for both THC-HF and THC-RPA once again demonstrates that all orbital blocks in ERIs are well described by THC. 
Even though the order of magnitude of the THC-HF and THC-RPA errors are different, we do see a systematic convergence trend in all quantities consistently. 
From the perspective of computational cost, in order to achieve 1 mHartree accuracy, our implementation of the THC-based algorithms (THC-HF/THC-RPA) are already faster than Chol-HF/Chol-RPA for our selected systems. 
As the number of $\textbf{k}$-points and the system size increase, the speedup in THC-HF and THC-RPA would be even more pronounced. 

\begin{figure}[tbh!]
\begin{center}
\includegraphics[width=0.45\textwidth]{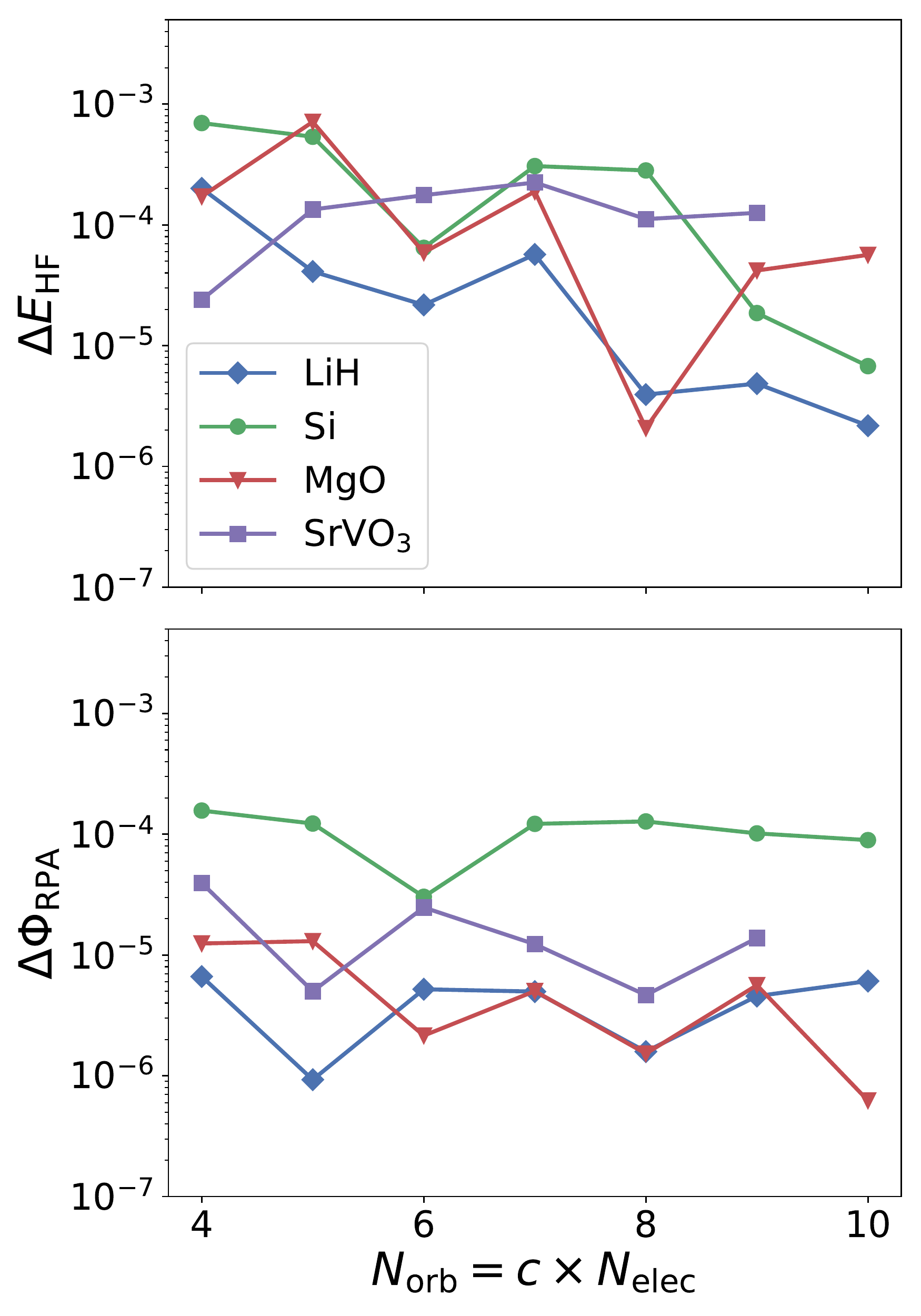}
\caption{Error of THC-HF and THC-RPA correlation energy per atom (Hartree) at $\alpha = 10$ with different numbers of basis functions $N_{\mathrm{orb}} = c N_{\mathrm{elec}}$. }
\label{fig:thc_rpa_basis_conv}
\end{center}
\end{figure}
Next, we analyze the error of the THC-based methods with respect to the number of single-particle basis functions. 
We construct the electronic Hamiltonians in THC factorization with $N_{\mathrm{orb}} = c N_{\mathrm{elec}}$ ($c=4\sim10$) at $\alpha_{\mathrm{THC}}=10$, and then perform THC-HF and THC-RPA calculations respectively. 
As shown in Fig.~\ref{fig:thc_rpa_basis_conv}, the accuracy of THC-HF and THC-RPA remains similarly as the size of the basis set increases, which suggests that the rank of ERI tensors scales linearly, rather than quadratically, with the number of basis functions. 
This behavior is observed in systems with different band gaps and even in a metallic system (SrVO$_{3}$) with transition metal atoms. 
Note that this is in contrast to Ref.\citenum{THC_MP3_Joonho2020} in which the error of THC-based methods is reported to increase as the size of the basis set enlarges.  
We believe the consistent accuracy observed in this work is due to the more robust choice of interpolating points provided by the pivoted Cholesky decomposition of the metric matrix, which leads to a consistent treatment of both occupied and virtual spaces. 
Such consistent accuracy among different physical systems manifest the power of THC-based methods compared to low-scaling algorithms which rely on sparsity and locality. 

\begin{figure}[tbh!]
\begin{center}
\includegraphics[width=0.45\textwidth]{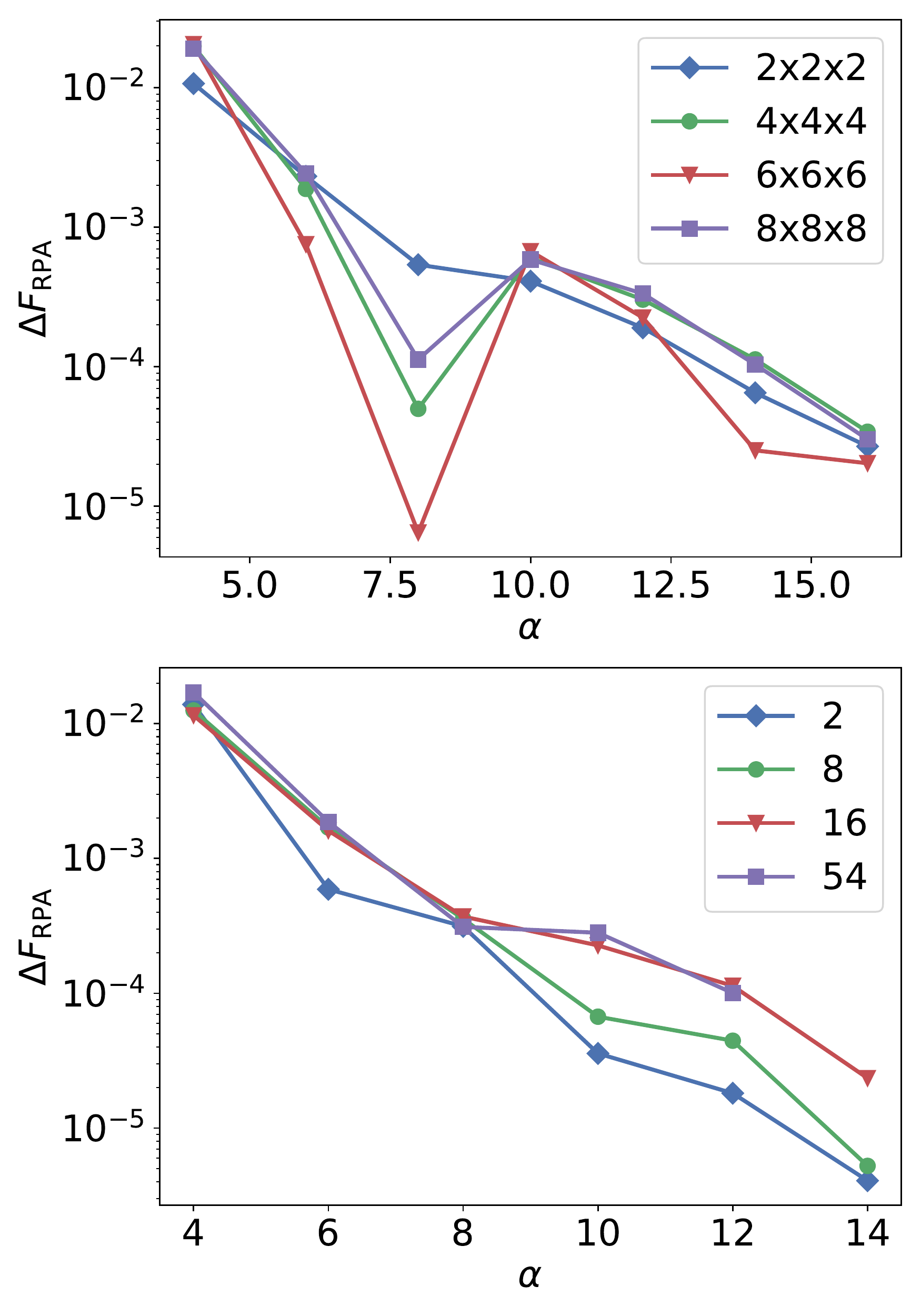}
\caption{Error of the RPA free energy per atom (Hartree) calculated using THC factorization as a function of $\alpha =  N_{\mathrm{\mu}}/N_{\mathrm{orb}}$: the primitive cell of Si on different $\textbf{k}$-meshes (top), Si supercells with different numbers of atoms (bottom). }
\label{fig:rpa_kpts_scell_conv}
\end{center}
\end{figure}
We now look at how the error of the THC decomposition behaves with respect to the number of $\textbf{k}$-points and the size of a supercell. 
As shown in Fig.~\ref{fig:rpa_kpts_scell_conv}, we compute the free energy in the random phase approximation using the THC decomposed ERIs for a primitive cell of Si on different $\textbf{k}$-meshes and $\Gamma$-point supercells of Si with different number of atoms. 
As we go to a larger $\textbf{k}$-mesh, the error of the free energy in the random phase approximation converges in a quantitatively similar manner. 
This is somewhat expected since the THC decomposition in our formulation is performed for each $\textbf{q}$-point independently, and therefore the $\textbf{q}$-dependent auxiliary bases are tailored to fit the Bloch pair densities for each $\textbf{q}$-point specifically. 
This further verifies that our previous analysis on a small $2\times2\times2$ $\textbf{k}$-mesh should be transferable to finer $\textbf{k}$-point sampling. 
Likewise, the error of RPA free energy per atom remains similarly as the size of the unit cell increases, especially for when $\alpha \leq 8$. 
This is consistent to Ref.\citenum{THC_MP3_Joonho2020} in which the error of extensive quantities scales linearly with the system size. 

\begin{figure}[tbh!]
\begin{center}
\includegraphics[width=0.45\textwidth]{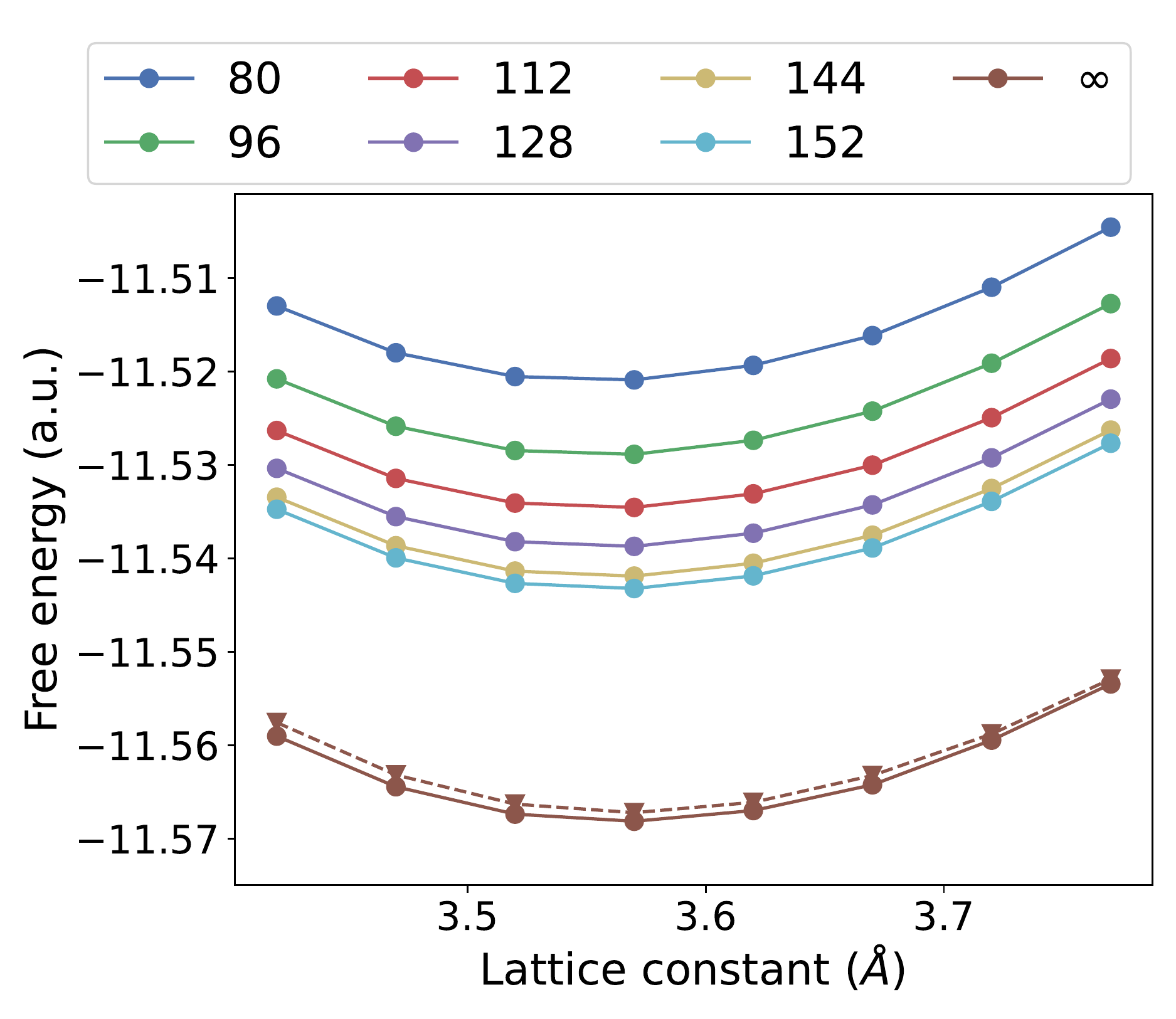}
\caption{Cold curves of diamond calculated from the THC-RPA free energy (solid lines) with different numbers of KS states. The infinite band limit is extrapolated from $\Phi^{\mathrm{RPA}}_{c}(N_{\mathrm{orb}}) = a/N_{\mathrm{orb}} + b$ as $N_{\mathrm{orb}}\rightarrow \infty$. The dashed line is calculated from \texttt{abinit}\cite{abinit_GONZE2020} and extrapolated using the same strategy. }
\label{fig:cold_curve}
\end{center}
\end{figure}
Lastly, we show the RPA equation of state of Carbon in the diamond phase in Fig.~\ref{fig:cold_curve}. 
The HF and the RPA correlation energy are calculated on a $15\times15\times15$ and a $8\times8\times8$ $\Gamma$-centered Monkhorst-Pack grid respectively. 
Due to the infinite summation over the virtual orbitals in the polarizability, the convergence of RPA correlation energy is notoriously slow with respect to the number of KS states\cite{RPA_Harl2008}. 
To obtain the converged values, we perform THC-RPA calculations for different numbers of KS orbitals and then extrapolated to the infinite basis set limit by fitting the formula $\Phi^{\mathrm{RPA}}_{c}(N_{\mathrm{orb}}) = a/N_{\mathrm{orb}} + b$.  
The Birch-Murnaghan equation\cite{Birch1947} is then fit to the extrapolated curve to extract the lattice constant ($a$) and bulk modulus ($B$). 
The predictions are $a = 3.57$ \AA \ and $B = 430$ GPa respectively. 
In addition, we have also performed RPA calculations using \texttt{abinit}\cite{abinit_GONZE2020} with the same numbers of KS orbitals and the same extrapolation strategy (dashed brown line). 
The results are $a=3.57$ \AA \ and $B=433$ GPa which is in a good agreement with our implementation. 

\subsection{Complexity analysis\label{subsec:complexity}}
\begin{figure}[tbh!]
\begin{center}
\includegraphics[width=0.55\textwidth]{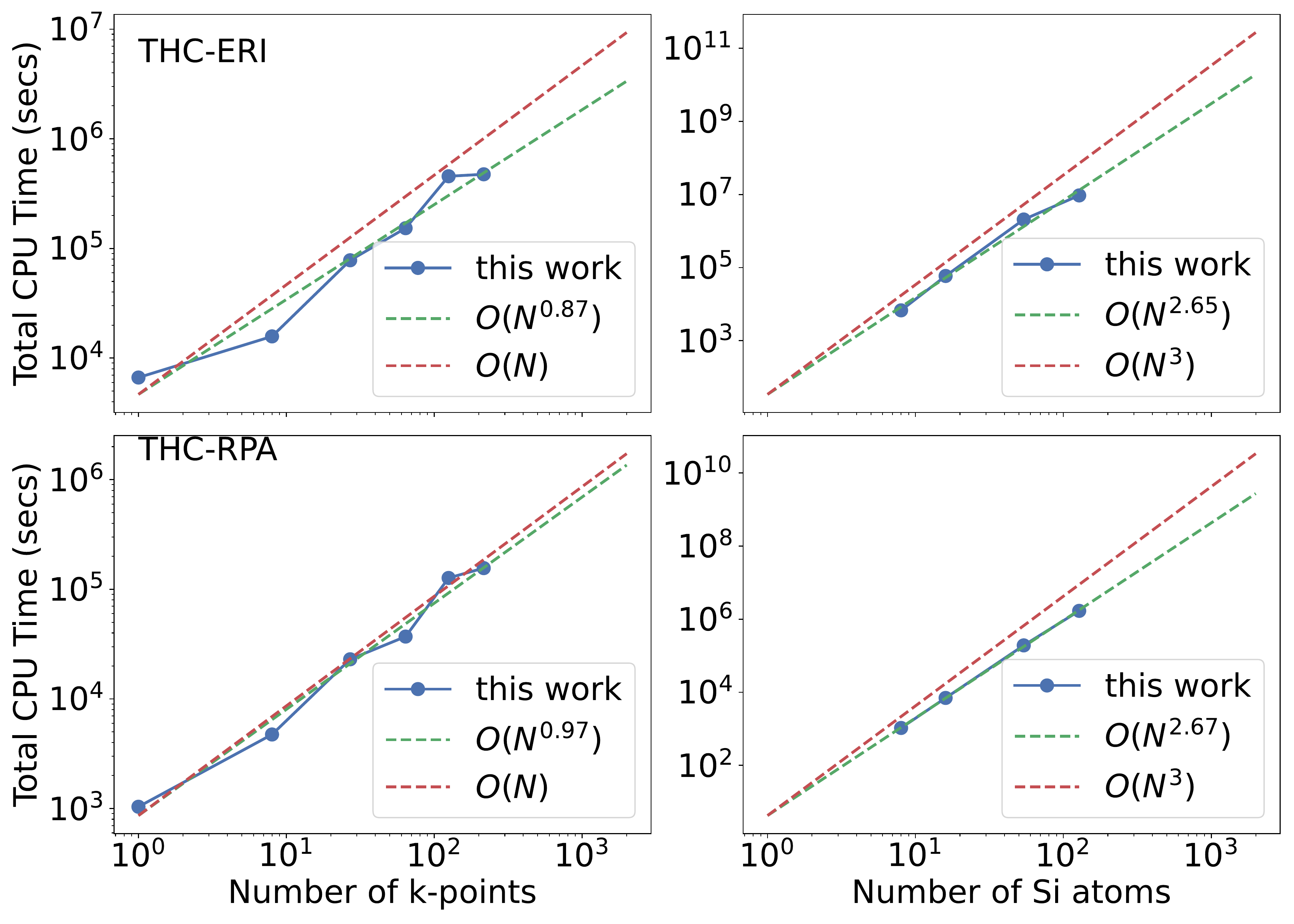}
\caption{Total CPU time for THC-ERI (first row) and THC-RPA (second row) with $\alpha = 10$. First column: the conventional unit cell of Si with different numbers of $\textbf{k}$-points. Second column: $\Gamma$-point Si supercells with increasing numbers of atoms.}
\label{fig:scaling}
\end{center}
\end{figure}
To demonstrate the low-scaling complexity of THC-based many-body perturbation theory, we show the total CPU timing of our THC-RPA implementations, including the steps for the preparation of ERI and the steps for the evaluation of RPA correlation energy. 
The systems are chosen to be the conventional unit cell of Si on a $n\times n \times n$ Monkhorst-Pack grid ($n=1\sim6$) and the $\Gamma$-point supercells of Si with 8, 16, 54, 128 atoms per unit cell. 
The kinetic energy cutoff is set to 30 hartree.  
As shown in Fig.~\ref{fig:scaling}, the time of the preparation of ERI and the steps for RPA energy scales linearly with the number of $\textbf{k}$-points and cubically with the number of atoms per unit cell. 
The observed speedup against to the $O(N_{k}N^{3})$ scaling is expected to be alleviated as the dimensions of a system further increase. 

Despite the same asymptotic scaling in the preparation and the RPA steps, the prefactors of these algorithms are quite different. 
In THC-ERI, the prefactor is proportional to $O(N_{\mu}^{2}N_{r})$ while the prefactor of the dominant steps in THC-RPA scales as $O(N_{\omega}N_{\mu}^{3})$, coming from Eq.~\ref{eq:phi_thc_rpa} where $N_{\omega}$ is the dimension of the Matsubara frequencies. 
Therefore, the relative computational cost of these two steps is given by the ratio of $N_{r}$ and $N_{\omega}N_{\mu}$. 
For the systems considered in this section, the timings of THC-ERI are slightly larger than those of THC-RPA. 
However, as the size of the single-particle basis increases, the cost of THC-RPA would increase faster compared to THC-ERI. 
In addition, THC-RPA becomes more expensive at lower temperature since the number of Matsubara frequency points required increases. 
On the other hand, for systems with very deep-lying orbitals, THC-ERI could become more computationally expensive due to a very large kinetic energy cutoff. 

\section{Conclusion \label{sec:conclusion}}
We introduce a low-scaling algorithm for RPA with $\textbf{k}$-point sampling based on the THC decomposition of ERIs. 
The THC representation of ERIs is achieved via a $\textbf{q}$-dependent ISDF procedure for Bloch pair densities in which both of the auxiliary basis and the fitting coefficients are computed on-the-fly for a given Bloch single-particle basis. 
Both the preparation steps of ERIs and the RPA parts scale linearly with the number of $\textbf{k}$-points and cubically with the system sizes due to the full separability of $\textbf{k}$-point and orbital indices in the THC representation of ERIs. 
The formalism is applicable to generic Bloch functions without an assumption on locality of orbitals, and its accuracy is systematically controlled by the size of the auxiliary basis. 
For our selected systems, we found $N_{\mu} = 8N_{\mathrm{orb}}$ is enough to achieve 1 mHartree accuracy for the ERI tensor, including all orbital blocks, and energies per atom. 
Such an observation is independent to the number of virtual orbitals, the number of $\textbf{k}$-points, and the size of a unit cell. 
The compactness of the size of the THC auxiliary basis enables many-body calculations for large-scale systems. 
Extending the periodic THC formulation to $GW$ and vertex corrections will be explored in the follow-up works, and the code will be made open-source in the near future.  

\begin{suppinfo}

Errors of ERI calculated from the THC factorization with \textbf{q}-dependent interpolating points; comparison with the original ISDF procedure for Bloch orbitals. 

\end{suppinfo}

\begin{acknowledgement}

We thank Alexander Hampel, Olivier Parcollet, and Antoine Georges for helpful discussions. We also thank Nils Wentzell for help with \texttt{nda} and \texttt{h5} libraries.
The Flatiron Institute is a division of the Simons Foundation.

\end{acknowledgement}

\bibliography{refs}

\providecommand{\latin}[1]{#1}
\makeatletter
\providecommand{\doi}
  {\begingroup\let\do\@makeother\dospecials
  \catcode`\{=1 \catcode`\}=2 \doi@aux}
\providecommand{\doi@aux}[1]{\endgroup\texttt{#1}}
\makeatother
\providecommand*\mcitethebibliography{\thebibliography}
\csname @ifundefined\endcsname{endmcitethebibliography}
  {\let\endmcitethebibliography\endthebibliography}{}
\begin{mcitethebibliography}{84}
\providecommand*\natexlab[1]{#1}
\providecommand*\mciteSetBstSublistMode[1]{}
\providecommand*\mciteSetBstMaxWidthForm[2]{}
\providecommand*\mciteBstWouldAddEndPuncttrue
  {\def\EndOfBibitem{\unskip.}}
\providecommand*\mciteBstWouldAddEndPunctfalse
  {\let\EndOfBibitem\relax}
\providecommand*\mciteSetBstMidEndSepPunct[3]{}
\providecommand*\mciteSetBstSublistLabelBeginEnd[3]{}
\providecommand*\EndOfBibitem{}
\mciteSetBstSublistMode{f}
\mciteSetBstMaxWidthForm{subitem}{(\alph{mcitesubitemcount})}
\mciteSetBstSublistLabelBeginEnd
  {\mcitemaxwidthsubitemform\space}
  {\relax}
  {\relax}

\bibitem[Hohenberg and Kohn(1964)Hohenberg, and Kohn]{Hohenberg1964}
Hohenberg,~P.; Kohn,~W. Inhomogeneous Electron Gas. \emph{Phys. Rev.}
  \textbf{1964}, \emph{136}, B864--B871\relax
\mciteBstWouldAddEndPuncttrue
\mciteSetBstMidEndSepPunct{\mcitedefaultmidpunct}
{\mcitedefaultendpunct}{\mcitedefaultseppunct}\relax
\EndOfBibitem
\bibitem[Kohn and Sham(1965)Kohn, and Sham]{KSDFT_Sham1965}
Kohn,~W.; Sham,~L.~J. Self-Consistent Equations Including Exchange and
  Correlation Effects. \emph{Phys. Rev.} \textbf{1965}, \emph{140},
  A1133--A1138\relax
\mciteBstWouldAddEndPuncttrue
\mciteSetBstMidEndSepPunct{\mcitedefaultmidpunct}
{\mcitedefaultendpunct}{\mcitedefaultseppunct}\relax
\EndOfBibitem
\bibitem[Medvedev \latin{et~al.}(2017)Medvedev, Bushmarinov, Sun, Perdew, and
  Lyssenko]{Michael2017}
Medvedev,~M.~G.; Bushmarinov,~I.~S.; Sun,~J.; Perdew,~J.~P.; Lyssenko,~K.~A.
  Density functional theory is straying from the path toward the exact
  functional. \emph{Science} \textbf{2017}, \emph{355}, 49--52\relax
\mciteBstWouldAddEndPuncttrue
\mciteSetBstMidEndSepPunct{\mcitedefaultmidpunct}
{\mcitedefaultendpunct}{\mcitedefaultseppunct}\relax
\EndOfBibitem
\bibitem[Onida \latin{et~al.}(2002)Onida, Reining, and
  Rubio]{DFTvsMBPT_Angel2002}
Onida,~G.; Reining,~L.; Rubio,~A. Electronic excitations: density-functional
  versus many-body Green's-function approaches. \emph{Rev. Mod. Phys.}
  \textbf{2002}, \emph{74}, 601--659\relax
\mciteBstWouldAddEndPuncttrue
\mciteSetBstMidEndSepPunct{\mcitedefaultmidpunct}
{\mcitedefaultendpunct}{\mcitedefaultseppunct}\relax
\EndOfBibitem
\bibitem[Langreth and Perdew(1977)Langreth, and
  Perdew]{Exc_jellium_Langreth1997}
Langreth,~D.~C.; Perdew,~J.~P. Exchange-correlation energy of a metallic
  surface: Wave-vector analysis. \emph{Phys. Rev. B} \textbf{1977}, \emph{15},
  2884--2901\relax
\mciteBstWouldAddEndPuncttrue
\mciteSetBstMidEndSepPunct{\mcitedefaultmidpunct}
{\mcitedefaultendpunct}{\mcitedefaultseppunct}\relax
\EndOfBibitem
\bibitem[Furche(2001)]{RPA_Furche2001}
Furche,~F. Molecular tests of the random phase approximation to the
  exchange-correlation energy functional. \emph{Phys. Rev. B} \textbf{2001},
  \emph{64}, 195120\relax
\mciteBstWouldAddEndPuncttrue
\mciteSetBstMidEndSepPunct{\mcitedefaultmidpunct}
{\mcitedefaultendpunct}{\mcitedefaultseppunct}\relax
\EndOfBibitem
\bibitem[Miyake \latin{et~al.}(2002)Miyake, Aryasetiawan, Kotani, van
  Schilfgaarde, Usuda, and Terakura]{RPA_Miyake2002}
Miyake,~T.; Aryasetiawan,~F.; Kotani,~T.; van Schilfgaarde,~M.; Usuda,~M.;
  Terakura,~K. Total energy of solids: An exchange and random-phase
  approximation correlation study. \emph{Phys. Rev. B} \textbf{2002},
  \emph{66}, 245103\relax
\mciteBstWouldAddEndPuncttrue
\mciteSetBstMidEndSepPunct{\mcitedefaultmidpunct}
{\mcitedefaultendpunct}{\mcitedefaultseppunct}\relax
\EndOfBibitem
\bibitem[Furche and Van~Voorhis(2005)Furche, and Van~Voorhis]{ACFDT_Furche2005}
Furche,~F.; Van~Voorhis,~T. Fluctuation-dissipation theorem density-functional
  theory. \emph{J. Chem. Phys.} \textbf{2005}, \emph{122}, 164106\relax
\mciteBstWouldAddEndPuncttrue
\mciteSetBstMidEndSepPunct{\mcitedefaultmidpunct}
{\mcitedefaultendpunct}{\mcitedefaultseppunct}\relax
\EndOfBibitem
\bibitem[Fuchs \latin{et~al.}(2005)Fuchs, Niquet, Gonze, and
  Burke]{Fuchs_RPA_2005}
Fuchs,~M.; Niquet,~Y.-M.; Gonze,~X.; Burke,~K. {Describing static correlation
  in bond dissociation by Kohn–Sham density functional theory}. \emph{J.
  Chem. Phys.} \textbf{2005}, \emph{122}, 094116\relax
\mciteBstWouldAddEndPuncttrue
\mciteSetBstMidEndSepPunct{\mcitedefaultmidpunct}
{\mcitedefaultendpunct}{\mcitedefaultseppunct}\relax
\EndOfBibitem
\bibitem[Marini \latin{et~al.}(2006)Marini, Garc\'{\i}a-Gonz\'alez, and
  Rubio]{ACFDT_hBN_Angel2006}
Marini,~A.; Garc\'{\i}a-Gonz\'alez,~P.; Rubio,~A. First-Principles Description
  of Correlation Effects in Layered Materials. \emph{Phys. Rev. Lett.}
  \textbf{2006}, \emph{96}, 136404\relax
\mciteBstWouldAddEndPuncttrue
\mciteSetBstMidEndSepPunct{\mcitedefaultmidpunct}
{\mcitedefaultendpunct}{\mcitedefaultseppunct}\relax
\EndOfBibitem
\bibitem[Furche(2008)]{ACFDT_RPA_Furche2008}
Furche,~F. Developing the random phase approximation into a practical
  post-Kohn–Sham correlation model. \emph{J. Chem. Phys.} \textbf{2008},
  \emph{129}, 114105\relax
\mciteBstWouldAddEndPuncttrue
\mciteSetBstMidEndSepPunct{\mcitedefaultmidpunct}
{\mcitedefaultendpunct}{\mcitedefaultseppunct}\relax
\EndOfBibitem
\bibitem[Ren \latin{et~al.}(2012)Ren, Rinke, Joas, and
  Scheffler]{RPA_review_Ren2012}
Ren,~X.; Rinke,~P.; Joas,~C.; Scheffler,~M. Random-phase approximation and its
  applications in computational chemistry and materials science. \emph{J.
  Mater. Sci.} \textbf{2012}, \emph{47}, 7447--7471\relax
\mciteBstWouldAddEndPuncttrue
\mciteSetBstMidEndSepPunct{\mcitedefaultmidpunct}
{\mcitedefaultendpunct}{\mcitedefaultseppunct}\relax
\EndOfBibitem
\bibitem[Grüneis \latin{et~al.}(2009)Grüneis, Marsman, Harl, Schimka, and
  Kresse]{SOSEX_Gruneis2009}
Grüneis,~A.; Marsman,~M.; Harl,~J.; Schimka,~L.; Kresse,~G. {Making the random
  phase approximation to electronic correlation accurate}. \emph{J. Chem.
  Phys.} \textbf{2009}, \emph{131}, 154115\relax
\mciteBstWouldAddEndPuncttrue
\mciteSetBstMidEndSepPunct{\mcitedefaultmidpunct}
{\mcitedefaultendpunct}{\mcitedefaultseppunct}\relax
\EndOfBibitem
\bibitem[Ángyán \latin{et~al.}(2011)Ángyán, Liu, Toulouse, and
  Jansen]{RPA_variants_Angyan2011}
Ángyán,~J.~G.; Liu,~R.-F.; Toulouse,~J.; Jansen,~G. Correlation Energy
  Expressions from the Adiabatic-Connection Fluctuation–Dissipation Theorem
  Approach. \emph{J. Chem. Theory Comput.} \textbf{2011}, \emph{7},
  3116--3130\relax
\mciteBstWouldAddEndPuncttrue
\mciteSetBstMidEndSepPunct{\mcitedefaultmidpunct}
{\mcitedefaultendpunct}{\mcitedefaultseppunct}\relax
\EndOfBibitem
\bibitem[Niquet \latin{et~al.}(2003)Niquet, Fuchs, and
  Gonze]{ACFDT_Niquet_2003}
Niquet,~Y.~M.; Fuchs,~M.; Gonze,~X. Exchange-correlation potentials in the
  adiabatic connection fluctuation-dissipation framework. \emph{Phys. Rev. A}
  \textbf{2003}, \emph{68}, 032507\relax
\mciteBstWouldAddEndPuncttrue
\mciteSetBstMidEndSepPunct{\mcitedefaultmidpunct}
{\mcitedefaultendpunct}{\mcitedefaultseppunct}\relax
\EndOfBibitem
\bibitem[Klein(1961)]{Klein_1961}
Klein,~A. Perturbation Theory for an Infinite Medium of Fermions. II.
  \emph{Phys. Rev.} \textbf{1961}, \emph{121}, 950--956\relax
\mciteBstWouldAddEndPuncttrue
\mciteSetBstMidEndSepPunct{\mcitedefaultmidpunct}
{\mcitedefaultendpunct}{\mcitedefaultseppunct}\relax
\EndOfBibitem
\bibitem[Dahlen \latin{et~al.}(2006)Dahlen, van Leeuwen, and von
  Barth]{Leeuwen_PRA_2006}
Dahlen,~N.~E.; van Leeuwen,~R.; von Barth,~U. Variational energy functionals of
  the Green function and of the density tested on molecules. \emph{Phys. Rev.
  A} \textbf{2006}, \emph{73}, 012511\relax
\mciteBstWouldAddEndPuncttrue
\mciteSetBstMidEndSepPunct{\mcitedefaultmidpunct}
{\mcitedefaultendpunct}{\mcitedefaultseppunct}\relax
\EndOfBibitem
\bibitem[Gell-Mann and Brueckner(1957)Gell-Mann, and
  Brueckner]{Corr_energy_eGas_GellMann1957}
Gell-Mann,~M.; Brueckner,~K.~A. Correlation Energy of an Electron Gas at High
  Density. \emph{Phys. Rev.} \textbf{1957}, \emph{106}, 364--368\relax
\mciteBstWouldAddEndPuncttrue
\mciteSetBstMidEndSepPunct{\mcitedefaultmidpunct}
{\mcitedefaultendpunct}{\mcitedefaultseppunct}\relax
\EndOfBibitem
\bibitem[Grüneis \latin{et~al.}(2010)Grüneis, Marsman, and
  Kresse]{MP2_solids_2010Gruneis}
Grüneis,~A.; Marsman,~M.; Kresse,~G. {Second-order Møller–Plesset
  perturbation theory applied to extended systems. II. Structural and energetic
  properties}. \emph{J. Chem. Phys.} \textbf{2010}, \emph{133}, 074107\relax
\mciteBstWouldAddEndPuncttrue
\mciteSetBstMidEndSepPunct{\mcitedefaultmidpunct}
{\mcitedefaultendpunct}{\mcitedefaultseppunct}\relax
\EndOfBibitem
\bibitem[Rojas \latin{et~al.}(1995)Rojas, Godby, and
  Needs]{spacetime_GW_Rojas1995}
Rojas,~H.~N.; Godby,~R.~W.; Needs,~R.~J. Space-Time Method for Ab Initio
  Calculations of Self-Energies and Dielectric Response Functions of Solids.
  \emph{Phys. Rev. Lett.} \textbf{1995}, \emph{74}, 1827--1830\relax
\mciteBstWouldAddEndPuncttrue
\mciteSetBstMidEndSepPunct{\mcitedefaultmidpunct}
{\mcitedefaultendpunct}{\mcitedefaultseppunct}\relax
\EndOfBibitem
\bibitem[Steinbeck \latin{et~al.}(2000)Steinbeck, Rubio, Reining, Torrent,
  White, and Godby]{spacetime_GW_enhancement_Steinbeck2000}
Steinbeck,~L.; Rubio,~A.; Reining,~L.; Torrent,~M.; White,~I.; Godby,~R.
  Enhancements to the GW space-time method. \emph{Comput. Phys. Commun.}
  \textbf{2000}, \emph{125}, 105--118\relax
\mciteBstWouldAddEndPuncttrue
\mciteSetBstMidEndSepPunct{\mcitedefaultmidpunct}
{\mcitedefaultendpunct}{\mcitedefaultseppunct}\relax
\EndOfBibitem
\bibitem[Bruneval and Gonze(2008)Bruneval, and Gonze]{GW_virtuals_Bruneval2008}
Bruneval,~F.; Gonze,~X. Accurate $GW$ self-energies in a plane-wave basis using
  only a few empty states: Towards large systems. \emph{Phys. Rev. B}
  \textbf{2008}, \emph{78}, 085125\relax
\mciteBstWouldAddEndPuncttrue
\mciteSetBstMidEndSepPunct{\mcitedefaultmidpunct}
{\mcitedefaultendpunct}{\mcitedefaultseppunct}\relax
\EndOfBibitem
\bibitem[Kaltak \latin{et~al.}(2014)Kaltak, Klime\ifmmode~\check{s}\else
  \v{s}\fi{}, and Kresse]{VASP_RPA_Kaltak2014}
Kaltak,~M.; Klime\ifmmode~\check{s}\else \v{s}\fi{},~J. c.~v.; Kresse,~G. Cubic
  scaling algorithm for the random phase approximation: Self-interstitials and
  vacancies in Si. \emph{Phys. Rev. B} \textbf{2014}, \emph{90}, 054115\relax
\mciteBstWouldAddEndPuncttrue
\mciteSetBstMidEndSepPunct{\mcitedefaultmidpunct}
{\mcitedefaultendpunct}{\mcitedefaultseppunct}\relax
\EndOfBibitem
\bibitem[Kaltak \latin{et~al.}(2014)Kaltak, Klimeš, and
  Kresse]{VASP_RPA_JCTC_Kaltak2014}
Kaltak,~M.; Klimeš,~J.; Kresse,~G. Low Scaling Algorithms for the Random Phase
  Approximation: Imaginary Time and Laplace Transformations. \emph{J. Chem.
  Theory Comput.} \textbf{2014}, \emph{10}, 2498--2507\relax
\mciteBstWouldAddEndPuncttrue
\mciteSetBstMidEndSepPunct{\mcitedefaultmidpunct}
{\mcitedefaultendpunct}{\mcitedefaultseppunct}\relax
\EndOfBibitem
\bibitem[Li \latin{et~al.}(2020)Li, Wallerberger, Chikano, Yeh, Gull, and
  Shinaoka]{sparse_sampling_Jia_2020}
Li,~J.; Wallerberger,~M.; Chikano,~N.; Yeh,~C.-N.; Gull,~E.; Shinaoka,~H.
  Sparse sampling approach to efficient ab initio calculations at finite
  temperature. \emph{Phys. Rev. B} \textbf{2020}, \emph{101}, 035144\relax
\mciteBstWouldAddEndPuncttrue
\mciteSetBstMidEndSepPunct{\mcitedefaultmidpunct}
{\mcitedefaultendpunct}{\mcitedefaultseppunct}\relax
\EndOfBibitem
\bibitem[Kaltak and Kresse(2020)Kaltak, and Kresse]{VASP_minimax_Kaltak2020}
Kaltak,~M.; Kresse,~G. Minimax isometry method: A compressive sensing approach
  for Matsubara summation in many-body perturbation theory. \emph{Phys. Rev. B}
  \textbf{2020}, \emph{101}, 205145\relax
\mciteBstWouldAddEndPuncttrue
\mciteSetBstMidEndSepPunct{\mcitedefaultmidpunct}
{\mcitedefaultendpunct}{\mcitedefaultseppunct}\relax
\EndOfBibitem
\bibitem[Gao \latin{et~al.}(2016)Gao, Xia, Gao, and Zhang]{GW_virutals_Gao2016}
Gao,~W.; Xia,~W.; Gao,~X.; Zhang,~P. Speeding up {GW} Calculations to Meet the
  Challenge of Large Scale Quasiparticle Predictions. \emph{Sci. Rep.}
  \textbf{2016}, \emph{6}\relax
\mciteBstWouldAddEndPuncttrue
\mciteSetBstMidEndSepPunct{\mcitedefaultmidpunct}
{\mcitedefaultendpunct}{\mcitedefaultseppunct}\relax
\EndOfBibitem
\bibitem[Kaye \latin{et~al.}(2022)Kaye, Chen, and Parcollet]{DLR_Kaye2022}
Kaye,~J.; Chen,~K.; Parcollet,~O. Discrete Lehmann representation of imaginary
  time Green's functions. \emph{Phys. Rev. B} \textbf{2022}, \emph{105},
  235115\relax
\mciteBstWouldAddEndPuncttrue
\mciteSetBstMidEndSepPunct{\mcitedefaultmidpunct}
{\mcitedefaultendpunct}{\mcitedefaultseppunct}\relax
\EndOfBibitem
\bibitem[Beebe and Linderberg(1977)Beebe, and
  Linderberg]{Cholesky_ERI_MOL_Beebe1977}
Beebe,~N. H.~F.; Linderberg,~J. Simplifications in the generation and
  transformation of two-electron integrals in molecular calculations.
  \emph{Int. J. Quantum Chem.} \textbf{1977}, \emph{12}, 683--705\relax
\mciteBstWouldAddEndPuncttrue
\mciteSetBstMidEndSepPunct{\mcitedefaultmidpunct}
{\mcitedefaultendpunct}{\mcitedefaultseppunct}\relax
\EndOfBibitem
\bibitem[Werner \latin{et~al.}(2003)Werner, Manby, and Knowles]{DF_Werner2003}
Werner,~H.-J.; Manby,~F.~R.; Knowles,~P.~J. Fast linear scaling second-order
  M{\o}ller-Plesset perturbation theory ({MP}2) using local and density fitting
  approximations. \emph{J. Chem. Phys.} \textbf{2003}, \emph{118},
  8149--8160\relax
\mciteBstWouldAddEndPuncttrue
\mciteSetBstMidEndSepPunct{\mcitedefaultmidpunct}
{\mcitedefaultendpunct}{\mcitedefaultseppunct}\relax
\EndOfBibitem
\bibitem[Ren \latin{et~al.}(2012)Ren, Rinke, Blum, Wieferink, Tkatchenko,
  Sanfilippo, Reuter, and Scheffler]{DF_Ren2012}
Ren,~X.; Rinke,~P.; Blum,~V.; Wieferink,~J.; Tkatchenko,~A.; Sanfilippo,~A.;
  Reuter,~K.; Scheffler,~M. Resolution-of-identity approach to
  Hartree{\textendash}Fock, hybrid density functionals, {RPA}, {MP}2
  {andGWwith} numeric atom-centered orbital basis functions. \emph{New J.
  Phys.} \textbf{2012}, \emph{14}, 053020\relax
\mciteBstWouldAddEndPuncttrue
\mciteSetBstMidEndSepPunct{\mcitedefaultmidpunct}
{\mcitedefaultendpunct}{\mcitedefaultseppunct}\relax
\EndOfBibitem
\bibitem[Sun \latin{et~al.}(2017)Sun, Berkelbach, McClain, and
  Chan]{GDF_MDF_Sun2017}
Sun,~Q.; Berkelbach,~T.~C.; McClain,~J.~D.; Chan,~G. K.-L. Gaussian and
  plane-wave mixed density fitting for periodic systems. \emph{J. Chem. Phys.}
  \textbf{2017}, \emph{147}, 164119\relax
\mciteBstWouldAddEndPuncttrue
\mciteSetBstMidEndSepPunct{\mcitedefaultmidpunct}
{\mcitedefaultendpunct}{\mcitedefaultseppunct}\relax
\EndOfBibitem
\bibitem[Ye and Berkelbach(2021)Ye, and Berkelbach]{RSDF_HongZhou2021}
Ye,~H.-Z.; Berkelbach,~T.~C. Fast periodic Gaussian density fitting by range
  separation. \emph{J. Chem. Phys.} \textbf{2021}, \emph{154}, 131104\relax
\mciteBstWouldAddEndPuncttrue
\mciteSetBstMidEndSepPunct{\mcitedefaultmidpunct}
{\mcitedefaultendpunct}{\mcitedefaultseppunct}\relax
\EndOfBibitem
\bibitem[Wilhelm \latin{et~al.}(2016)Wilhelm, Seewald, Del~Ben, and
  Hutter]{CP2K_cubicRPA_2016}
Wilhelm,~J.; Seewald,~P.; Del~Ben,~M.; Hutter,~J. Large-Scale Cubic-Scaling
  Random Phase Approximation Correlation Energy Calculations Using a Gaussian
  Basis. \emph{J. Chem. Theory Comput.} \textbf{2016}, \emph{12},
  5851--5859\relax
\mciteBstWouldAddEndPuncttrue
\mciteSetBstMidEndSepPunct{\mcitedefaultmidpunct}
{\mcitedefaultendpunct}{\mcitedefaultseppunct}\relax
\EndOfBibitem
\bibitem[Schurkus and Ochsenfeld(2016)Schurkus, and
  Ochsenfeld]{lineaRPA_AO_Schurkus2016}
Schurkus,~H.~F.; Ochsenfeld,~C. {Communication: An effective linear-scaling
  atomic-orbital reformulation of the random-phase approximation using a
  contracted double-Laplace transformation}. \emph{J. Chem. Phys.}
  \textbf{2016}, \emph{144}, 031101\relax
\mciteBstWouldAddEndPuncttrue
\mciteSetBstMidEndSepPunct{\mcitedefaultmidpunct}
{\mcitedefaultendpunct}{\mcitedefaultseppunct}\relax
\EndOfBibitem
\bibitem[Luenser \latin{et~al.}(2017)Luenser, Schurkus, and
  Ochsenfeld]{linearRPA_Luenser2017}
Luenser,~A.; Schurkus,~H.~F.; Ochsenfeld,~C. Vanishing-Overhead Linear-Scaling
  Random Phase Approximation by Cholesky Decomposition and an Attenuated
  Coulomb-Metric. \emph{J. Chem. Theory Comput.} \textbf{2017}, \emph{13},
  1647--1655\relax
\mciteBstWouldAddEndPuncttrue
\mciteSetBstMidEndSepPunct{\mcitedefaultmidpunct}
{\mcitedefaultendpunct}{\mcitedefaultseppunct}\relax
\EndOfBibitem
\bibitem[Kállay(2015)]{linearDRPA_Kallay2015}
Kállay,~M. {Linear-scaling implementation of the direct random-phase
  approximation}. \emph{J. Chem. Phys.} \textbf{2015}, \emph{142}, 204105\relax
\mciteBstWouldAddEndPuncttrue
\mciteSetBstMidEndSepPunct{\mcitedefaultmidpunct}
{\mcitedefaultendpunct}{\mcitedefaultseppunct}\relax
\EndOfBibitem
\bibitem[Morales and Malone(2020)Morales, and Malone]{GTO_AFQMC_Miguel2020}
Morales,~M.~A.; Malone,~F.~D. {Accelerating the convergence of auxiliary-field
  quantum Monte Carlo in solids with optimized Gaussian basis sets}. \emph{J.
  Chem. Phys.} \textbf{2020}, \emph{153}, 194111\relax
\mciteBstWouldAddEndPuncttrue
\mciteSetBstMidEndSepPunct{\mcitedefaultmidpunct}
{\mcitedefaultendpunct}{\mcitedefaultseppunct}\relax
\EndOfBibitem
\bibitem[Zhou \latin{et~al.}(2021)Zhou, Gull, and Zgid]{GTO_Zhou2021}
Zhou,~Y.; Gull,~E.; Zgid,~D. Material-Specific Optimization of Gaussian Basis
  Sets against Plane Wave Data. \emph{J. Chem. Theory Comput.} \textbf{2021},
  \emph{17}, 5611--5622\relax
\mciteBstWouldAddEndPuncttrue
\mciteSetBstMidEndSepPunct{\mcitedefaultmidpunct}
{\mcitedefaultendpunct}{\mcitedefaultseppunct}\relax
\EndOfBibitem
\bibitem[Ye and Berkelbach(2022)Ye, and Berkelbach]{ccgto_Ye2022}
Ye,~H.-Z.; Berkelbach,~T.~C. Correlation-Consistent Gaussian Basis Sets for
  Solids Made Simple. \emph{J. Chem. Theory Comput.} \textbf{2022}, \emph{18},
  1595--1606\relax
\mciteBstWouldAddEndPuncttrue
\mciteSetBstMidEndSepPunct{\mcitedefaultmidpunct}
{\mcitedefaultendpunct}{\mcitedefaultseppunct}\relax
\EndOfBibitem
\bibitem[Hohenstein \latin{et~al.}(2012)Hohenstein, Parrish, and
  Martínez]{THC_one_Martinez2012}
Hohenstein,~E.~G.; Parrish,~R.~M.; Martínez,~T.~J. {Tensor hypercontraction
  density fitting. I. Quartic scaling second- and third-order Møller-Plesset
  perturbation theory}. \emph{J. Chem. Phys.} \textbf{2012}, \emph{137},
  044103\relax
\mciteBstWouldAddEndPuncttrue
\mciteSetBstMidEndSepPunct{\mcitedefaultmidpunct}
{\mcitedefaultendpunct}{\mcitedefaultseppunct}\relax
\EndOfBibitem
\bibitem[Parrish \latin{et~al.}(2012)Parrish, Hohenstein, Martínez, and
  Sherrill]{LSTHC_Sherrill2012}
Parrish,~R.~M.; Hohenstein,~E.~G.; Martínez,~T.~J.; Sherrill,~C.~D. Tensor
  hypercontraction. II. Least-squares renormalization. \emph{J. Chem. Phys.}
  \textbf{2012}, \emph{137}, 224106\relax
\mciteBstWouldAddEndPuncttrue
\mciteSetBstMidEndSepPunct{\mcitedefaultmidpunct}
{\mcitedefaultendpunct}{\mcitedefaultseppunct}\relax
\EndOfBibitem
\bibitem[Parrish \latin{et~al.}(2013)Parrish, Hohenstein, Martínez, and
  Sherrill]{THC_DVR_Sherrill2013}
Parrish,~R.~M.; Hohenstein,~E.~G.; Martínez,~T.~J.; Sherrill,~C.~D. {Discrete
  variable representation in electronic structure theory: Quadrature grids for
  least-squares tensor hypercontraction}. \emph{J. Chem. Phys.} \textbf{2013},
  \emph{138}, 194107\relax
\mciteBstWouldAddEndPuncttrue
\mciteSetBstMidEndSepPunct{\mcitedefaultmidpunct}
{\mcitedefaultendpunct}{\mcitedefaultseppunct}\relax
\EndOfBibitem
\bibitem[Kokkila~Schumacher \latin{et~al.}(2015)Kokkila~Schumacher, Hohenstein,
  Parrish, Wang, and Martínez]{LSTHC_MP2_Schumacher2015}
Kokkila~Schumacher,~S. I.~L.; Hohenstein,~E.~G.; Parrish,~R.~M.; Wang,~L.-P.;
  Martínez,~T.~J. Tensor Hypercontraction Second-Order Møller–Plesset
  Perturbation Theory: Grid Optimization and Reaction Energies. \emph{J. Chem.
  Theory Comput.} \textbf{2015}, \emph{11}, 3042--3052\relax
\mciteBstWouldAddEndPuncttrue
\mciteSetBstMidEndSepPunct{\mcitedefaultmidpunct}
{\mcitedefaultendpunct}{\mcitedefaultseppunct}\relax
\EndOfBibitem
\bibitem[Lu and Ying(2015)Lu, and Ying]{ISDF_Lu2015}
Lu,~J.; Ying,~L. Compression of the electron repulsion integral tensor in
  tensor hypercontraction format with cubic scaling cost. \emph{J. Comput.
  Phys.} \textbf{2015}, \emph{302}, 329--335\relax
\mciteBstWouldAddEndPuncttrue
\mciteSetBstMidEndSepPunct{\mcitedefaultmidpunct}
{\mcitedefaultendpunct}{\mcitedefaultseppunct}\relax
\EndOfBibitem
\bibitem[Lu and Ying(2016)Lu, and Ying]{ISDF_Bloch_Lu2016}
Lu,~J.; Ying,~L. Fast algorithm for periodic density fitting for Bloch waves.
  \emph{Annals of Mathematical Sciences and Applications} \textbf{2016},
  \emph{1}, 321--339\relax
\mciteBstWouldAddEndPuncttrue
\mciteSetBstMidEndSepPunct{\mcitedefaultmidpunct}
{\mcitedefaultendpunct}{\mcitedefaultseppunct}\relax
\EndOfBibitem
\bibitem[Hu \latin{et~al.}(2017)Hu, Lin, and Yang]{ISDF_QRCP_hybrid_Hu2017}
Hu,~W.; Lin,~L.; Yang,~C. Interpolative Separable Density Fitting Decomposition
  for Accelerating Hybrid Density Functional Calculations with Applications to
  Defects in Silicon. \emph{J. Chem. Theory Comput.} \textbf{2017}, \emph{13},
  5420--5431\relax
\mciteBstWouldAddEndPuncttrue
\mciteSetBstMidEndSepPunct{\mcitedefaultmidpunct}
{\mcitedefaultendpunct}{\mcitedefaultseppunct}\relax
\EndOfBibitem
\bibitem[Dong \latin{et~al.}(2018)Dong, Hu, and Lin]{ISDF_CVT_hybrid_Lin2018}
Dong,~K.; Hu,~W.; Lin,~L. Interpolative Separable Density Fitting through
  Centroidal Voronoi Tessellation with Applications to Hybrid Functional
  Electronic Structure Calculations. \emph{J. Chem. Theory Comput.}
  \textbf{2018}, \emph{14}, 1311--1320\relax
\mciteBstWouldAddEndPuncttrue
\mciteSetBstMidEndSepPunct{\mcitedefaultmidpunct}
{\mcitedefaultendpunct}{\mcitedefaultseppunct}\relax
\EndOfBibitem
\bibitem[Qin \latin{et~al.}(2020)Qin, Liu, Hu, and
  Yang]{ISDF_hybridDFT_NAO_Qin2020}
Qin,~X.; Liu,~J.; Hu,~W.; Yang,~J. Interpolative Separable Density Fitting
  Decomposition for Accelerating Hartree–Fock Exchange Calculations within
  Numerical Atomic Orbitals. \emph{J. Phys. Chem. A} \textbf{2020}, \emph{124},
  5664--5674\relax
\mciteBstWouldAddEndPuncttrue
\mciteSetBstMidEndSepPunct{\mcitedefaultmidpunct}
{\mcitedefaultendpunct}{\mcitedefaultseppunct}\relax
\EndOfBibitem
\bibitem[Qin \latin{et~al.}(2020)Qin, Li, Hu, and
  Yang]{ISDF_CVT_hybrid_NAO_Qin2020}
Qin,~X.; Li,~J.; Hu,~W.; Yang,~J. Machine Learning K-Means Clustering Algorithm
  for Interpolative Separable Density Fitting to Accelerate Hybrid Functional
  Calculations with Numerical Atomic Orbitals. \emph{J. Phys. Chem. A}
  \textbf{2020}, \emph{124}, 10066--10074\relax
\mciteBstWouldAddEndPuncttrue
\mciteSetBstMidEndSepPunct{\mcitedefaultmidpunct}
{\mcitedefaultendpunct}{\mcitedefaultseppunct}\relax
\EndOfBibitem
\bibitem[Sharma \latin{et~al.}(2022)Sharma, White, and
  Beylkin]{rPS_HF_Sharma2022}
Sharma,~S.; White,~A.~F.; Beylkin,~G. Fast Exchange with Gaussian Basis Set
  Using Robust Pseudospectral Method. \emph{J. Chem. Theory Comput.}
  \textbf{2022}, \emph{18}, 7306--7320\relax
\mciteBstWouldAddEndPuncttrue
\mciteSetBstMidEndSepPunct{\mcitedefaultmidpunct}
{\mcitedefaultendpunct}{\mcitedefaultseppunct}\relax
\EndOfBibitem
\bibitem[Hohenstein \latin{et~al.}(2012)Hohenstein, Parrish, Sherrill, and
  Martínez]{THC_three_Martinez2012}
Hohenstein,~E.~G.; Parrish,~R.~M.; Sherrill,~C.~D.; Martínez,~T.~J.
  {Communication: Tensor hypercontraction. III. Least-squares tensor
  hypercontraction for the determination of correlated wavefunctions}. \emph{J.
  Chem. Phys.} \textbf{2012}, \emph{137}, 221101\relax
\mciteBstWouldAddEndPuncttrue
\mciteSetBstMidEndSepPunct{\mcitedefaultmidpunct}
{\mcitedefaultendpunct}{\mcitedefaultseppunct}\relax
\EndOfBibitem
\bibitem[Hohenstein \latin{et~al.}(2013)Hohenstein, Kokkila, Parrish, and
  Martínez]{THC_CC2_Hohenstein2013}
Hohenstein,~E.~G.; Kokkila,~S. I.~L.; Parrish,~R.~M.; Martínez,~T.~J. {Quartic
  scaling second-order approximate coupled cluster singles and doubles via
  tensor hypercontraction: THC-CC2}. \emph{J. Chem. Phys.} \textbf{2013},
  \emph{138}, 124111\relax
\mciteBstWouldAddEndPuncttrue
\mciteSetBstMidEndSepPunct{\mcitedefaultmidpunct}
{\mcitedefaultendpunct}{\mcitedefaultseppunct}\relax
\EndOfBibitem
\bibitem[Hohenstein \latin{et~al.}(2013)Hohenstein, Kokkila, Parrish, and
  Martínez]{THC_EOMCC2_Hohenstein2013}
Hohenstein,~E.~G.; Kokkila,~S. I.~L.; Parrish,~R.~M.; Martínez,~T.~J. Tensor
  Hypercontraction Equation-of-Motion Second-Order Approximate Coupled Cluster:
  Electronic Excitation Energies in O(N4) Time. \emph{J. Phys. Chem. B}
  \textbf{2013}, \emph{117}, 12972--12978\relax
\mciteBstWouldAddEndPuncttrue
\mciteSetBstMidEndSepPunct{\mcitedefaultmidpunct}
{\mcitedefaultendpunct}{\mcitedefaultseppunct}\relax
\EndOfBibitem
\bibitem[Parrish \latin{et~al.}(2014)Parrish, Sherrill, Hohenstein, Kokkila,
  and Martínez]{LSTHC_CCSD_Parrish2014}
Parrish,~R.~M.; Sherrill,~C.~D.; Hohenstein,~E.~G.; Kokkila,~S. I.~L.;
  Martínez,~T.~J. {Communication: Acceleration of coupled cluster singles and
  doubles via orbital-weighted least-squares tensor hypercontraction}. \emph{J.
  Chem. Phys.} \textbf{2014}, \emph{140}, 181102\relax
\mciteBstWouldAddEndPuncttrue
\mciteSetBstMidEndSepPunct{\mcitedefaultmidpunct}
{\mcitedefaultendpunct}{\mcitedefaultseppunct}\relax
\EndOfBibitem
\bibitem[Song and Martínez(2016)Song, and Martínez]{THC_SOS_MP2_one_Song2016}
Song,~C.; Martínez,~T.~J. {Atomic orbital-based SOS-MP2 with tensor
  hypercontraction. I. GPU-based tensor construction and exploiting sparsity}.
  \emph{J. Chem. Phys.} \textbf{2016}, \emph{144}, 174111\relax
\mciteBstWouldAddEndPuncttrue
\mciteSetBstMidEndSepPunct{\mcitedefaultmidpunct}
{\mcitedefaultendpunct}{\mcitedefaultseppunct}\relax
\EndOfBibitem
\bibitem[Song and Martínez(2017)Song, and Martínez]{THC_SOS_MP2_two_Song2017}
Song,~C.; Martínez,~T.~J. {Atomic orbital-based SOS-MP2 with tensor
  hypercontraction. II. Local tensor hypercontraction}. \emph{J. Chem. Phys.}
  \textbf{2017}, \emph{146}, 034104\relax
\mciteBstWouldAddEndPuncttrue
\mciteSetBstMidEndSepPunct{\mcitedefaultmidpunct}
{\mcitedefaultendpunct}{\mcitedefaultseppunct}\relax
\EndOfBibitem
\bibitem[Song and Martínez(2017)Song, and
  Martínez]{THC_SOS_MP2_gradient_Song2017}
Song,~C.; Martínez,~T.~J. {Analytical gradients for tensor hyper-contracted
  MP2 and SOS-MP2 on graphical processing units}. \emph{J. Chem. Phys.}
  \textbf{2017}, \emph{147}, 161723\relax
\mciteBstWouldAddEndPuncttrue
\mciteSetBstMidEndSepPunct{\mcitedefaultmidpunct}
{\mcitedefaultendpunct}{\mcitedefaultseppunct}\relax
\EndOfBibitem
\bibitem[Lee \latin{et~al.}(2020)Lee, Lin, and Head-Gordon]{THC_MP3_Joonho2020}
Lee,~J.; Lin,~L.; Head-Gordon,~M. Systematically Improvable Tensor
  Hypercontraction: Interpolative Separable Density-Fitting for Molecules
  Applied to Exact Exchange, Second- and Third-Order Møller–Plesset
  Perturbation Theory. \emph{J. Chem. Theory Comput.} \textbf{2020}, \emph{16},
  243--263\relax
\mciteBstWouldAddEndPuncttrue
\mciteSetBstMidEndSepPunct{\mcitedefaultmidpunct}
{\mcitedefaultendpunct}{\mcitedefaultseppunct}\relax
\EndOfBibitem
\bibitem[Lu and Thicke(2017)Lu, and Thicke]{ISDF_RPA_model_Lu2017}
Lu,~J.; Thicke,~K. Cubic scaling algorithms for RPA correlation using
  interpolative separable density fitting. \emph{J. Comput. Phys.}
  \textbf{2017}, \emph{351}, 187--202\relax
\mciteBstWouldAddEndPuncttrue
\mciteSetBstMidEndSepPunct{\mcitedefaultmidpunct}
{\mcitedefaultendpunct}{\mcitedefaultseppunct}\relax
\EndOfBibitem
\bibitem[Duchemin and Blase(2019)Duchemin, and
  Blase]{Separable_RI_RPA_Blase2019}
Duchemin,~I.; Blase,~X. Separable resolution-of-the-identity with all-electron
  Gaussian bases: Application to cubic-scaling RPA. \emph{J. Chem. Phys.}
  \textbf{2019}, \emph{150}, 174120\relax
\mciteBstWouldAddEndPuncttrue
\mciteSetBstMidEndSepPunct{\mcitedefaultmidpunct}
{\mcitedefaultendpunct}{\mcitedefaultseppunct}\relax
\EndOfBibitem
\bibitem[Gao and Chelikowsky(2020)Gao, and
  Chelikowsky]{TDDFT_G0W0_ISDF_Gao2020}
Gao,~W.; Chelikowsky,~J.~R. Accelerating Time-Dependent Density Functional
  Theory and GW Calculations for Molecules and Nanoclusters with Symmetry
  Adapted Interpolative Separable Density Fitting. \emph{J. Chem. Theory
  Comput.} \textbf{2020}, \emph{16}, 2216--2223\relax
\mciteBstWouldAddEndPuncttrue
\mciteSetBstMidEndSepPunct{\mcitedefaultmidpunct}
{\mcitedefaultendpunct}{\mcitedefaultseppunct}\relax
\EndOfBibitem
\bibitem[Ma \latin{et~al.}(2021)Ma, Wang, Wan, Li, Qin, Liu, Hu, Lin, Yang, and
  Yang]{G0W0_COHSEX_ISDF_Ma2021}
Ma,~H.; Wang,~L.; Wan,~L.; Li,~J.; Qin,~X.; Liu,~J.; Hu,~W.; Lin,~L.; Yang,~C.;
  Yang,~J. Realizing Effective Cubic-Scaling Coulomb Hole Plus Screened
  Exchange Approximation in Periodic Systems via Interpolative Separable
  Density Fitting with a Plane-Wave Basis Set. \emph{J. Phys. Chem. A}
  \textbf{2021}, \emph{125}, 7545--7557\relax
\mciteBstWouldAddEndPuncttrue
\mciteSetBstMidEndSepPunct{\mcitedefaultmidpunct}
{\mcitedefaultendpunct}{\mcitedefaultseppunct}\relax
\EndOfBibitem
\bibitem[Duchemin and Blase(2021)Duchemin, and
  Blase]{separable_RI_G0W0_Blase2021}
Duchemin,~I.; Blase,~X. Cubic-Scaling All-Electron GW Calculations with a
  Separable Density-Fitting Space–Time Approach. \emph{J. Chem. Theory
  Comput.} \textbf{2021}, \emph{17}, 2383--2393\relax
\mciteBstWouldAddEndPuncttrue
\mciteSetBstMidEndSepPunct{\mcitedefaultmidpunct}
{\mcitedefaultendpunct}{\mcitedefaultseppunct}\relax
\EndOfBibitem
\bibitem[Malone \latin{et~al.}(2019)Malone, Zhang, and
  Morales]{AFQMC_ISDF_Miguel2019}
Malone,~F.~D.; Zhang,~S.; Morales,~M.~A. Overcoming the Memory Bottleneck in
  Auxiliary Field Quantum Monte Carlo Simulations with Interpolative Separable
  Density Fitting. \emph{J. Chem. Theory Comput.} \textbf{2019}, \emph{15},
  256--264\relax
\mciteBstWouldAddEndPuncttrue
\mciteSetBstMidEndSepPunct{\mcitedefaultmidpunct}
{\mcitedefaultendpunct}{\mcitedefaultseppunct}\relax
\EndOfBibitem
\bibitem[Wu \latin{et~al.}(2022)Wu, Qin, Hu, and
  Yang]{ISDF_hybridDFT_kpts_2022}
Wu,~K.; Qin,~X.; Hu,~W.; Yang,~J. Low-Rank Approximations Accelerated
  Plane-Wave Hybrid Functional Calculations with k-Point Sampling. \emph{J.
  Chem. Theory Comput.} \textbf{2022}, \emph{18}, 206--218\relax
\mciteBstWouldAddEndPuncttrue
\mciteSetBstMidEndSepPunct{\mcitedefaultmidpunct}
{\mcitedefaultendpunct}{\mcitedefaultseppunct}\relax
\EndOfBibitem
\bibitem[Cheng \latin{et~al.}(2005)Cheng, Gimbutas, Martinsson, and
  Rokhlin]{ID_Cheng2005}
Cheng,~H.; Gimbutas,~Z.; Martinsson,~P.~G.; Rokhlin,~V. On the Compression of
  Low Rank Matrices. \emph{SIAM J. Sci. Comput.} \textbf{2005}, \emph{26},
  1389--1404\relax
\mciteBstWouldAddEndPuncttrue
\mciteSetBstMidEndSepPunct{\mcitedefaultmidpunct}
{\mcitedefaultendpunct}{\mcitedefaultseppunct}\relax
\EndOfBibitem
\bibitem[Liberty \latin{et~al.}(2007)Liberty, Woolfe, Martinsson, Rokhlin, and
  Tygert]{ID_Liberty2007}
Liberty,~E.; Woolfe,~F.; Martinsson,~P.-G.; Rokhlin,~V.; Tygert,~M. Randomized
  algorithms for the low-rank approximation of matrices. \emph{Proc. Natl.
  Acad. Sci.} \textbf{2007}, \emph{104}, 20167--20172\relax
\mciteBstWouldAddEndPuncttrue
\mciteSetBstMidEndSepPunct{\mcitedefaultmidpunct}
{\mcitedefaultendpunct}{\mcitedefaultseppunct}\relax
\EndOfBibitem
\bibitem[Matthews(2020)]{LSTHC_Matthews2020}
Matthews,~D.~A. Improved Grid Optimization and Fitting in Least Squares Tensor
  Hypercontraction. \emph{J. Chem. Theory Comput.} \textbf{2020}, \emph{16},
  1382--1385\relax
\mciteBstWouldAddEndPuncttrue
\mciteSetBstMidEndSepPunct{\mcitedefaultmidpunct}
{\mcitedefaultendpunct}{\mcitedefaultseppunct}\relax
\EndOfBibitem
\bibitem[Luttinger and Ward(1960)Luttinger, and Ward]{LW_1960}
Luttinger,~J.~M.; Ward,~J.~C. Ground-State Energy of a Many-Fermion System. II.
  \emph{Phys. Rev.} \textbf{1960}, \emph{118}, 1417--1427\relax
\mciteBstWouldAddEndPuncttrue
\mciteSetBstMidEndSepPunct{\mcitedefaultmidpunct}
{\mcitedefaultendpunct}{\mcitedefaultseppunct}\relax
\EndOfBibitem
\bibitem[Perdew \latin{et~al.}(1996)Perdew, Burke, and
  Ernzerhof]{PBE_Perdew1996}
Perdew,~J.~P.; Burke,~K.; Ernzerhof,~M. Generalized Gradient Approximation Made
  Simple. \emph{Phys. Rev. Lett.} \textbf{1996}, \emph{77}, 3865--3868\relax
\mciteBstWouldAddEndPuncttrue
\mciteSetBstMidEndSepPunct{\mcitedefaultmidpunct}
{\mcitedefaultendpunct}{\mcitedefaultseppunct}\relax
\EndOfBibitem
\bibitem[Giannozzi \latin{et~al.}(2009)Giannozzi, Baroni, Bonini, Calandra,
  Car, Cavazzoni, Ceresoli, Chiarotti, Cococcioni, Dabo, Corso, de~Gironcoli,
  Fabris, Fratesi, Gebauer, Gerstmann, Gougoussis, Kokalj, Lazzeri,
  Martin-Samos, Marzari, Mauri, Mazzarello, Paolini, Pasquarello, Paulatto,
  Sbraccia, Scandolo, Sclauzero, Seitsonen, Smogunov, Umari, and
  Wentzcovitch]{QE_Giannozzi2009}
Giannozzi,~P.; Baroni,~S.; Bonini,~N.; Calandra,~M.; Car,~R.; Cavazzoni,~C.;
  Ceresoli,~D.; Chiarotti,~G.~L.; Cococcioni,~M.; Dabo,~I.; Corso,~A.~D.;
  de~Gironcoli,~S.; Fabris,~S.; Fratesi,~G.; Gebauer,~R.; Gerstmann,~U.;
  Gougoussis,~C.; Kokalj,~A.; Lazzeri,~M.; Martin-Samos,~L.; Marzari,~N.;
  Mauri,~F.; Mazzarello,~R.; Paolini,~S.; Pasquarello,~A.; Paulatto,~L.;
  Sbraccia,~C.; Scandolo,~S.; Sclauzero,~G.; Seitsonen,~A.~P.; Smogunov,~A.;
  Umari,~P.; Wentzcovitch,~R.~M. QUANTUM ESPRESSO: a modular and open-source
  software project for quantum simulations of materials. \emph{J. Phys.:
  Condens. Matter} \textbf{2009}, \emph{21}, 395502\relax
\mciteBstWouldAddEndPuncttrue
\mciteSetBstMidEndSepPunct{\mcitedefaultmidpunct}
{\mcitedefaultendpunct}{\mcitedefaultseppunct}\relax
\EndOfBibitem
\bibitem[Giannozzi \latin{et~al.}(2017)Giannozzi, Andreussi, Brumme, Bunau,
  Nardelli, Calandra, Car, Cavazzoni, Ceresoli, Cococcioni, Colonna, Carnimeo,
  Corso, de~Gironcoli, Delugas, DiStasio, Ferretti, Floris, Fratesi, Fugallo,
  Gebauer, Gerstmann, Giustino, Gorni, Jia, Kawamura, Ko, Kokalj,
  Küçükbenli, Lazzeri, Marsili, Marzari, Mauri, Nguyen, Nguyen, de-la Roza,
  Paulatto, Poncé, Rocca, Sabatini, Santra, Schlipf, Seitsonen, Smogunov,
  Timrov, Thonhauser, Umari, Vast, Wu, and Baroni]{QE_Giannozzi2017}
Giannozzi,~P.; Andreussi,~O.; Brumme,~T.; Bunau,~O.; Nardelli,~M.~B.;
  Calandra,~M.; Car,~R.; Cavazzoni,~C.; Ceresoli,~D.; Cococcioni,~M.;
  Colonna,~N.; Carnimeo,~I.; Corso,~A.~D.; de~Gironcoli,~S.; Delugas,~P.;
  DiStasio,~R.~A.; Ferretti,~A.; Floris,~A.; Fratesi,~G.; Fugallo,~G.;
  Gebauer,~R.; Gerstmann,~U.; Giustino,~F.; Gorni,~T.; Jia,~J.; Kawamura,~M.;
  Ko,~H.-Y.; Kokalj,~A.; Küçükbenli,~E.; Lazzeri,~M.; Marsili,~M.;
  Marzari,~N.; Mauri,~F.; Nguyen,~N.~L.; Nguyen,~H.-V.; de-la Roza,~A.~O.;
  Paulatto,~L.; Poncé,~S.; Rocca,~D.; Sabatini,~R.; Santra,~B.; Schlipf,~M.;
  Seitsonen,~A.~P.; Smogunov,~A.; Timrov,~I.; Thonhauser,~T.; Umari,~P.;
  Vast,~N.; Wu,~X.; Baroni,~S. Advanced capabilities for materials modelling
  with Quantum ESPRESSO. \emph{J. Phys.: Condens. Matter} \textbf{2017},
  \emph{29}, 465901\relax
\mciteBstWouldAddEndPuncttrue
\mciteSetBstMidEndSepPunct{\mcitedefaultmidpunct}
{\mcitedefaultendpunct}{\mcitedefaultseppunct}\relax
\EndOfBibitem
\bibitem[Giannozzi \latin{et~al.}(2020)Giannozzi, Baseggio, Bonf{\`{a}},
  Brunato, Car, Carnimeo, Cavazzoni, de~Gironcoli, Delugas, Ruffino, Ferretti,
  Marzari, Timrov, Urru, and Baroni]{QE_Giannozzi2020}
Giannozzi,~P.; Baseggio,~O.; Bonf{\`{a}},~P.; Brunato,~D.; Car,~R.;
  Carnimeo,~I.; Cavazzoni,~C.; de~Gironcoli,~S.; Delugas,~P.; Ruffino,~F.~F.;
  Ferretti,~A.; Marzari,~N.; Timrov,~I.; Urru,~A.; Baroni,~S. Quantum ESPRESSO
  toward the exascale. \emph{J. Chem. Phys.} \textbf{2020}, \emph{152},
  154105\relax
\mciteBstWouldAddEndPuncttrue
\mciteSetBstMidEndSepPunct{\mcitedefaultmidpunct}
{\mcitedefaultendpunct}{\mcitedefaultseppunct}\relax
\EndOfBibitem
\bibitem[Hamann(2013)]{ONCVPP_Hamann2013}
Hamann,~D.~R. Optimized norm-conserving Vanderbilt pseudopotentials.
  \emph{Phys. Rev. B} \textbf{2013}, \emph{88}, 085117\relax
\mciteBstWouldAddEndPuncttrue
\mciteSetBstMidEndSepPunct{\mcitedefaultmidpunct}
{\mcitedefaultendpunct}{\mcitedefaultseppunct}\relax
\EndOfBibitem
\bibitem[Schlipf and Gygi(2015)Schlipf, and Gygi]{SG15ONCV_Schlipf2015}
Schlipf,~M.; Gygi,~F. Optimization algorithm for the generation of ONCV
  pseudopotentials. \emph{Comput. Phys. Commun.} \textbf{2015}, \emph{196},
  36--44\relax
\mciteBstWouldAddEndPuncttrue
\mciteSetBstMidEndSepPunct{\mcitedefaultmidpunct}
{\mcitedefaultendpunct}{\mcitedefaultseppunct}\relax
\EndOfBibitem
\bibitem[{van Setten} \latin{et~al.}(2018){van Setten}, Giantomassi, Bousquet,
  Verstraete, Hamann, Gonze, and Rignanese]{pseudodojo_2018}
{van Setten},~M.; Giantomassi,~M.; Bousquet,~E.; Verstraete,~M.; Hamann,~D.;
  Gonze,~X.; Rignanese,~G.-M. The PseudoDojo: Training and grading a 85 element
  optimized norm-conserving pseudopotential table. \emph{Comput. Phys. Commun.}
  \textbf{2018}, \emph{226}, 39--54\relax
\mciteBstWouldAddEndPuncttrue
\mciteSetBstMidEndSepPunct{\mcitedefaultmidpunct}
{\mcitedefaultendpunct}{\mcitedefaultseppunct}\relax
\EndOfBibitem
\bibitem[Shinaoka \latin{et~al.}(2017)Shinaoka, Otsuki, Ohzeki, and
  Yoshimi]{IR_Hiroshi_2017}
Shinaoka,~H.; Otsuki,~J.; Ohzeki,~M.; Yoshimi,~K. Compressing Green's function
  using intermediate representation between imaginary-time and real-frequency
  domains. \emph{Phys. Rev. B} \textbf{2017}, \emph{96}, 035147\relax
\mciteBstWouldAddEndPuncttrue
\mciteSetBstMidEndSepPunct{\mcitedefaultmidpunct}
{\mcitedefaultendpunct}{\mcitedefaultseppunct}\relax
\EndOfBibitem
\bibitem[Wallerberger \latin{et~al.}(2023)Wallerberger, Badr, Hoshino, Huber,
  Kakizawa, Koretsune, Nagai, Nogaki, Nomoto, Mori, Otsuki, Ozaki, Plaikner,
  Sakurai, Vogel, Witt, Yoshimi, and Shinaoka]{spare_ir_Markus2023}
Wallerberger,~M.; Badr,~S.; Hoshino,~S.; Huber,~S.; Kakizawa,~F.;
  Koretsune,~T.; Nagai,~Y.; Nogaki,~K.; Nomoto,~T.; Mori,~H.; Otsuki,~J.;
  Ozaki,~S.; Plaikner,~T.; Sakurai,~R.; Vogel,~C.; Witt,~N.; Yoshimi,~K.;
  Shinaoka,~H. sparse-ir: Optimal compression and sparse sampling of many-body
  propagators. \emph{SoftwareX} \textbf{2023}, \emph{21}, 101266\relax
\mciteBstWouldAddEndPuncttrue
\mciteSetBstMidEndSepPunct{\mcitedefaultmidpunct}
{\mcitedefaultendpunct}{\mcitedefaultseppunct}\relax
\EndOfBibitem
\bibitem[Yeh \latin{et~al.}(2022)Yeh, Iskakov, Zgid, and Gull]{scGW_CNY2022}
Yeh,~C.-N.; Iskakov,~S.; Zgid,~D.; Gull,~E. Fully self-consistent
  finite-temperature $GW$ in Gaussian Bloch orbitals for solids. \emph{Phys.
  Rev. B} \textbf{2022}, \emph{106}, 235104\relax
\mciteBstWouldAddEndPuncttrue
\mciteSetBstMidEndSepPunct{\mcitedefaultmidpunct}
{\mcitedefaultendpunct}{\mcitedefaultseppunct}\relax
\EndOfBibitem
\bibitem[Gonze \latin{et~al.}(2020)Gonze, Amadon, Antonius, Arnardi, Baguet,
  Beuken, Bieder, Bottin, Bouchet, Bousquet, Brouwer, Bruneval, Brunin,
  Cavignac, Charraud, Chen, Côté, Cottenier, Denier, Geneste, Ghosez,
  Giantomassi, Gillet, Gingras, Hamann, Hautier, He, Helbig, Holzwarth, Jia,
  Jollet, Lafargue-Dit-Hauret, Lejaeghere, Marques, Martin, Martins, Miranda,
  Naccarato, Persson, Petretto, Planes, Pouillon, Prokhorenko, Ricci,
  Rignanese, Romero, Schmitt, Torrent, {van Setten}, {Van Troeye}, Verstraete,
  Zérah, and Zwanziger]{abinit_GONZE2020}
Gonze,~X.; Amadon,~B.; Antonius,~G.; Arnardi,~F.; Baguet,~L.; Beuken,~J.-M.;
  Bieder,~J.; Bottin,~F.; Bouchet,~J.; Bousquet,~E.; Brouwer,~N.; Bruneval,~F.;
  Brunin,~G.; Cavignac,~T.; Charraud,~J.-B.; Chen,~W.; Côté,~M.;
  Cottenier,~S.; Denier,~J.; Geneste,~G.; Ghosez,~P.; Giantomassi,~M.;
  Gillet,~Y.; Gingras,~O.; Hamann,~D.~R.; Hautier,~G.; He,~X.; Helbig,~N.;
  Holzwarth,~N.; Jia,~Y.; Jollet,~F.; Lafargue-Dit-Hauret,~W.; Lejaeghere,~K.;
  Marques,~M.~A.; Martin,~A.; Martins,~C.; Miranda,~H.~P.; Naccarato,~F.;
  Persson,~K.; Petretto,~G.; Planes,~V.; Pouillon,~Y.; Prokhorenko,~S.;
  Ricci,~F.; Rignanese,~G.-M.; Romero,~A.~H.; Schmitt,~M.~M.; Torrent,~M.; {van
  Setten},~M.~J.; {Van Troeye},~B.; Verstraete,~M.~J.; Zérah,~G.;
  Zwanziger,~J.~W. The Abinitproject: Impact, environment and recent
  developments. \emph{Comput. Phys. Commun.} \textbf{2020}, \emph{248},
  107042\relax
\mciteBstWouldAddEndPuncttrue
\mciteSetBstMidEndSepPunct{\mcitedefaultmidpunct}
{\mcitedefaultendpunct}{\mcitedefaultseppunct}\relax
\EndOfBibitem
\bibitem[Harl and Kresse(2008)Harl, and Kresse]{RPA_Harl2008}
Harl,~J.; Kresse,~G. Cohesive energy curves for noble gas solids calculated by
  adiabatic connection fluctuation-dissipation theory. \emph{Phys. Rev. B}
  \textbf{2008}, \emph{77}, 045136\relax
\mciteBstWouldAddEndPuncttrue
\mciteSetBstMidEndSepPunct{\mcitedefaultmidpunct}
{\mcitedefaultendpunct}{\mcitedefaultseppunct}\relax
\EndOfBibitem
\bibitem[Birch(1947)]{Birch1947}
Birch,~F. Finite Elastic Strain of Cubic Crystals. \emph{Phys. Rev.}
  \textbf{1947}, \emph{71}, 809--824\relax
\mciteBstWouldAddEndPuncttrue
\mciteSetBstMidEndSepPunct{\mcitedefaultmidpunct}
{\mcitedefaultendpunct}{\mcitedefaultseppunct}\relax
\EndOfBibitem
\end{mcitethebibliography}

\end{document}